\documentclass[aps,pra,reprint,showpacs,showkeys,groupedaddress,nofootinbib,superscriptaddress]{revtex4-1}
\usepackage{hyperref}
\usepackage[T1]{fontenc}
\usepackage[ansinew]{inputenc}
\usepackage{amsmath}
\usepackage{setspace}
\usepackage{psfrag}
\usepackage{mathtools}
\usepackage{amssymb,lineno,amsfonts}
\usepackage{verbatim}
\usepackage{color}
\usepackage{textcomp}
\usepackage[toc,page]{appendix}
\usepackage{graphicx}
\usepackage[caption=false]{subfig}
\usepackage{float}
\usepackage{bm}
\usepackage{dcolumn}
\usepackage{gensymb}
\usepackage{array}
\usepackage{ifpdf}
\usepackage{bbm}
\usepackage{mathrsfs}
\usepackage{upgreek}
\usepackage{epstopdf}
\usepackage{esvect}
\usepackage[usenames,dvipsnames]{xcolor}
\definecolor{med-blue}{RGB}{25,25,112}
\hypersetup{colorlinks, linkcolor={blue},citecolor={blue}, urlcolor={blue}}
\usepackage[english]{babel}
\usepackage[autostyle, english = american]{csquotes}
\usepackage{enumerate}
\usepackage{amsfonts}
\usepackage{amssymb}
\begin{document}

\title{A variational approach for the ground state profile of a trapped spinor-BEC: A detailed study of phase transition in spin-1 condensate at zero magnetic field }

\author{Projjwal K. Kanjilal}
\email{projjwal.kanjilal@students.iiserpune.ac.in}
 \affiliation{Indian Institute of Science Education and Research,Pune}
\author{A. Bhattacharyay}
\email{a.bhattacharyay@iiserpune.ac.in}
\affiliation{Indian Institute of Science Education and Research,Pune}

\date{\today}

\begin{abstract}{In this article we introduce a multi-modal variational method to analytically estimate the full number- and corresponding energy-density profile of a spin-1 Bose-Einstein condensate (BEC) for number of particles as low as 500 under harmonic confinement. To apply this method, we consider a system of spin-1 BEC under three dimensional isotropic and effective one-dimensional harmonic confinement in absence (negligible presence) of magnetic field which has ground state candidates of comparable energy. It should be noted that in such circumstances kinetic energy contribution to the ground state cannot be neglected which puts the applicability of Thomas-Fermi approximation to question. For anti-ferromagnetic condensates, the T-F approximated energy difference between the competing stationary states (ground state and the first excited state) is approximately 0.3\%. As T-F approximation is only good for condensates with large number of particles, T-F approximated predictions can completely go wrong especially for small condensates. This is where comes the role of a detailed analysis using our variational method, which incorporates the kinetic energy contribution and accurately estimates the number- and energy-density profile even for condensates having small number of particles. Results of our analytical method are supported by numerical simulation. This variational method is a general and can be extended to other similar/higher-dimensional problems to get results beyond the accuracy of Thomas-Fermi approximation.}
    



\end{abstract}

\maketitle


\section{Introduction}
With the advent of optical trapping technique the spin degrees of freedom of a trapped Bose-Einstein condensate (BEC) become accessible degrees of freedom \cite{PhysRevLett.80.2027}. Such a system is often commonly called as spinor BEC and due to its complexity and rich ground state structures \cite{Zhang_2003,mats10gr,KAWAGUCHI2012253,stenger98,PhysRevLett.81.5718} it has gained a lot of momentum in both analytical, computational and experimental studies.

By applying small magnetic field one can manipulate the population of different spin components by tuning linear and quadratic Zeeman terms, thus producing rich phase diagrams characteristic to the system. This easy tunability of magnetic terms paved the way of growing interest of exploring phase transition \cite{computational}, phase separation and domain formation of different stationary phases \cite{PhysRevA.85.023601,PhysRevA.84.013619,PhysRevA.96.043603,PhysRevA.60.4857,PhysRevA.78.023632,PhysRevA.80.023602,PhysRevLett.77.3276,2399-6528-2-2-025008,PhysRevA.85.043602,doi:10.1063/1.3243875,doi:10.1142/S0217984917502153,PhysRevA.82.033609,li2018phase,PhysRevA.94.013602,sabbatini2011phase,gautam2014phase,gautam2011phase,PhysRevA.72.023610,Kanjilal_2020}. Many associated phenomena arise including excitation, instabilities and associated quasi-particles across phase boundaries \cite{vidanovic2013spin,PhysRevResearch.3.023043}.

Many of these theoretical and computational studies of ground state structure hinges upon investigating the role of spin-spin interactions on the outcome in terms of an ordered configuration. The investigation of the role of such interactions require a negligible presence of external field.  As a result there has been a lot of analysis on the ground state when the quadratic and linear Zeeman terms are zero \cite{PhysRevA.92.023616,jimenez2018spontaneous}. 

In the seminal work by T.-L. Ho \cite{PhysRevLett.81.742} the ground state structure in a harmonic confinement was investigated and it was shown that in the ferromagnetic type of condensate the ground state structure is ferromagnetic type and for anti-ferromagnetic interaction the ground state is polar state. It was also argued that due to spin rotational symmetry, different degenerate states are equally likely to be the ground state. It is obvious that the anti-phase-matched (APM) state (where all the three spin-components are populated with relative phase being $\pi$), polar (where only the zeroth spin component is populated) and the anti-ferromagnetic state falls in this degeneracy. 

In the contrast to T.-L. Ho \cite{PhysRevLett.81.742}, in the article \cite{PhysRevA.66.011601}, the focus is on single mode approximation (SMA) where, all the sub-components share the same spatial profile. The ground state structure is investigated with three-component states (3-C) to see the validity of SMA. Due to fixing of magnetization ($M\neq 1$) phase-matched (PM) state (where, all the spin components are populated with relative phase being zero) is found as ground state which follows SMA (for ferromagnetic type of interaction) but in anti-ferromagnetic condensate APM state, which does not follow SMA, becomes the ground state.
\par
Another interesting theoretical approach is to start with non-equilibrium populations in different spin components and with time evolution the equilibrium densities in ground state be achieved \cite{PhysRevA.60.1463}. W. Zhang $et$ $al.$ \cite{Zhang_2003} studied the ground state structure both in harmonic and box confinement in fixed magnetization. This changes the ground state structure due to different magnetization. It has been shown that only at zero magnetic field, $B=0$ and zero magnetization, $M=0$ the single mode approximation is valid, otherwise not in general this method holds in the case of anti-ferromagnetic type of condensate.

In article \cite{jimenez2018spontaneous} the ground state structures are seen in zero magnetic force. Under conservation of the total magnetization, domain formation by different stationary states is also reported.
The phase diagram in a trapped condensate can also be obtained via modulational instability for anti-ferromagnetic condensate \cite{PhysRevA.78.023632}.
The ground state \cite{PhysRevA.92.023616} in non-zero and zero magnetization is also seen using SMA in ferromagnetic type condensate. The ground state phase diagram of a spin-1 BEC is shown to get substantially modified due to finite temperature effects \cite{PhysRevA.84.043645,PhysRevA.85.053611} and interparticle correlations \cite{PhysRevA.102.013302}. 
\par
In this paper, we present a general multi-modal variational method based on matching of the profiles in the higher and lower density regions which would work for even small condensates of a few hundred particles. This variational approach was partly developed in \cite{Pendse2017} by Pendse and Bhattacharyay for non-spinor BEC. This multi-modal approach, in contrast to SMA, considers the number densities for individual Zeeman components. Corresponding to the smooth density profile of the Zeeman-components of the condensate, this variational method ensures a smooth energy-density profile where the kenetic-energy is not neglected like what happens in a T-F approximation. This is the reason that the present variational method provides one with an almost exact number- and energy-density for the spinor condensate even for small number of particles. The present method makes use of the knowledge of the first few harmonic oscillator states to determine the tail part of the distribution which is completely neglected in a T-F approximation. Moreover, this redistribution of the particles of the condensate also substantially alters the core density profile from that which is obtained using T-F approximation. 

It is well known that, the T-F approximation is a very good one for larger condensates, however, there often appear marginal cases of comparison of energy of the phases under T-F approximation which require a more rigorous treatment. In the present context, we investigate  exactly such a problem related to the ground state of a trapped spin-1 BEC under the influence of spin-spin interactions when the external magnetic fields are negligible (practically not present). We will show in this paper that the results obtained on the basis of T-F approximation are indeed marginal in the presence of energetically competing stationary states which are ground state candidates. A substantial resolution of the problem requires a better method which takes into account the kinetic energy of the system which is neglected in the T-F approximation. Our variational method works exceptionally well in such circumstances. Our method provides a detailed and rigorous insight into the problem at hand which is beyond the scope of T-F approximation. Our method is general enough to be applied to other scenarios of concern where T-F approximation has reasons to be considered inconclusive.

The paper is organized in the following way. In section-II we introduce the system and present the set of all allowed stationary phases. In section-III the phase boundary against the variation of the spin-spin interaction parameter $c_1$ is determined based on T-F approximation. In section-IV we consider the full profile of the condensate with the introduction of the variational method. We carry out the detailed calculations to identify the corrections and observe the quantitative effect of the corrections on total energies, energy density and particle number density of the stationary states. Following that we conclude the paper with a discussion.

\section{Mean field description for spin-1 condensate}

Spin-1 BEC in mean-field regime is well described by a set of three coupled Gross-Pitaevskii equations, \cite{PhysRevLett.81.742,doi:10.1143/JPSJ.67.1822,Kanjilal_2020}
\begin{equation}\label{eq:rev1}
    i \hbar\dfrac{\partial \psi_m}{\partial t}= \left(\mathcal{H}-pm + qm^2\right)\psi_m
    +c_1 \sum_{m'=-1}^1 \vec{F}.\vec{f}_{mm'} \psi_{m'},
\end{equation} 
where, $\psi_m$ is the mean field corresponding to the $m^{th}$ sub-component which also serves as an order parameter of the system. Here, the suffix $m$ and $m'$ run from $-1$ to $1$ in integer steps for a spin-1 BEC. $\mathcal{H}$ is a short-hand notation for the symmetric part of the Hamiltonian, $\mathcal{H}=-\dfrac{\hbar^2 \nabla^2}{2M} + U(\vec{r})+c_0 n$, where $c_0$ is the spin-independent interaction coefficient and $U(\vec{r})$ is the trapping potential. The effect of magnetic field comes into the dynamics via the linear and quadratic Zeeman terms $p$ and $q$ respectively.
$c_1$ in the last term of Eq.(\ref{eq:rev1}) represents the spin-dependent interaction coefficient and $\vec{f}$ denotes the spin-1 matrices, \cite{KAWAGUCHI2012253}
{\small
	$$
	f_x=\frac{1}{\sqrt{2}}
	\begin{bmatrix}
	0 & 1 & 0 \\
	1 & 0 & 1 \\
	0 & 1 & 0
	\end{bmatrix},
	f_y=\frac{i}{\sqrt{2}}
	\begin{bmatrix}-
	0 & -1 & 0 \\
	1 & 0 & -1 \\
	0 & 1 & 0
	\end{bmatrix},
	f_z=
	\begin{bmatrix}
	1 & 0 & 0 \\
	0 & 0 & 0 \\
	0 & 0 & -1  
	\end{bmatrix}.
	$$}
The local spin density $\vec{F}$ is defined as,
\begin{equation}\label{eq:rev2}
    \vec{F}=\sum^{1}_{i,j=-1} \psi^{*}_i \vec{f}_{ij}\psi_j.
\end{equation}

\par

The mean-field total energy, \cite{KAWAGUCHI2012253}
\begin{equation}\label{eq:rev3}
    \begin{split}
        E=\int d\vec{r} \sum^1_{m=-1} \psi_m^* &\left( -\dfrac{\hbar^2 \nabla^2}{2M} + U_{trap}(\vec{r}) -pm + qm^2  \right)\psi_m\\
        &\qquad\qquad\qquad+ \dfrac{c_0}{2} n^2+\dfrac{c_1}{2} |\vec{F}|^2,
    \end{split}
\end{equation}
serves as a guiding parameter, comparison of which for the stationary states reveals the ground state structures.
In this paper the stationary states are represented as $(\mathbbm{n}_1,\mathbbm{n}_0,\mathbbm{n}_{-1})$, where, if the sub-components are (un-) populated we represent it as binary notation, (0 or) 1. For example to represent the anti-ferromagnetic state we stick to the notation $(1,0,1)$, which tells that the magnetic sub-levels, $m_{F}=1$ and $-1$ are populated and the zeroth sub-level is empty.

\subsection{Dynamics of sub-component density and phases}
 For a trapped spin-1 BEC under an arbitrary trapping potential $U(\vec{r})$, the GP equation can be simplified by plugging in the ansatz,
 \begin{equation}\label{eq:rev4}
    \psi_m(\vec{r},t)=\sqrt{n_m(\vec{r})}exp(-\dfrac{i\mu t}{\hbar})exp(-i\theta_m).
 \end{equation}
 This allows to get the density and phase dynamics separately for all the sub-components as, \cite{Kanjilal_2020} 
 \begin{equation}\label{eq:rev5}
    \dot{n}_0(\vec{r})=-\dfrac{4c_1 n_0 \sqrt{n_1n_{-1}}\sin\theta_r}{\hbar},
 \end{equation}
 \begin{equation}\label{eq:rev6}
    \dot{n}_{\pm1}(\vec{r})=\dfrac{2c_1 n_0 \sqrt{n_1n_{-1}}\sin\theta_r}{\hbar},
 \end{equation}
 \begin{equation}\label{eq:rev7}
    \begin{split}
        \hbar\dot{\theta}_0=\dfrac{1}{\sqrt{n_0(\vec{r})}}&\left(\mathcal{H}-\mu\right)\sqrt{n_0(\vec{r})}\\
        &+c_1\left(n_1+n_{-1}+2 \sqrt{n_{-1}n_1}\cos\theta_r\right),
    \end{split}
 \end{equation}
 \begin{equation}\label{eq:rev8}
    \begin{split}
        \hbar\dot{\theta}_{\pm1}=\dfrac{1}{\sqrt{n_{\pm1}(\vec{r})}}&\left(\mathcal{H}-\mu\right)\sqrt{n_{\pm1}(\vec{r})}\pm c_1\left(n_1-n_{-1}\right)+q\\
        &\quad \mp p+c_1n_0\left(1 +\sqrt{\dfrac{n_{\mp1}(\vec{r})}{n_{\pm1}(\vec{r})}}\cos\theta_r\right),
    \end{split}
 \end{equation}
  where, the relative phase is defined as, $\theta_r=\theta_1+\theta_{-1}-2\theta_0$ \cite{KAWAGUCHI2012253}.
 For stationary states, the density and phases of each sub-component should not vary with time, so one can replace the left hand side of Eqs.(\ref{eq:rev5}-\ref{eq:rev8}) with zero. In this article, we are not interested in the vortex solutions, so the individual component phases are treated as spatially invariant or the gradient of the phases is negligible.
 The energy density (the integrand in Eq.(\ref{eq:rev3})) thus, can be written in simplified way following the ansatz (Eq.(\ref{eq:rev4})),
 \begin{equation}\label{eq:rev9}
    \begin{split}
        e(\vec{r})=-\sum^1_{m=-1} &\sqrt{n_m(\vec{r})}\dfrac{\hbar^2 \nabla^2}{2M}\sqrt{n_m(\vec{r})}+U(\vec{r})n(\vec{r})\\ &+\dfrac{c_0}{2}n^2(\vec{r})+\dfrac{c_1}{2}\left(n_1-n_{-1}\right)^2\\
        &+ \dfrac{c_1}{2}n_0\left[n_1+n_{-1}+2\sqrt{n_1n_{-1}}cos\theta_r\right].
    \end{split} 
 \end{equation}
 
 From the stationarity condition (viz. $\dot{n}_m=0$) in Eq.(\ref{eq:rev5}-\ref{eq:rev6}), it is evident that either (at least) one of the sub-components has to be empty or all the sub-components are populated provided that the condition, $\sin{\theta_r} =0$ is satisfied. So, when all the sub-components are populated the stationary state corresponding to $\theta_r=0$ is the phase-matched (PM) state and $\theta_r=\pi$ is the anti-phase-matched (APM) state. The stationary states, for which, only one of the sub-components is populated, are the two ferromagnetic states ($(1,0,0)$ and $(0,0,1)$) and the polar state, $(0,1,0)$. Whereas, we get three possible mixed states when only one of the sub-components is empty. This includes two possible mixed-ferromagnetic states ($(1,1,0)$ and $(0,1,1)$) and the anti-ferromagnetic state, $(1,0,1)$. 
 \par
 Solutions of the Eqs.(\ref{eq:rev5}-\ref{eq:rev6}) following the stationarity conditions yield different stationary state structures. For different stationary states, the solution of the Eqs.(\ref{eq:rev7}-\ref{eq:rev8}) provides the number- and corresponding energy-density profile \cite{Kanjilal_2020}, which depend on the Zeeman terms ($p$, $q$), chemical potential, trapping potential and the interaction terms ($c_0$ and $c_1$).
 
 \subsection{Stationary states at p=0 and q=0 using T-F approximation}
    It is interesting to observe the physics at zero magnetic field where the linear and quadratic Zeeman term, denoted as $p$ and $q$ respectively, vanish \cite{PhysRevLett.81.742,PhysRevA.60.1463,PhysRevA.92.023616}.

    Stationary states existing at this limit can be found by putting $p=0$ and $q=0$ in Eq.(\ref{eq:rev8}). The number density and energy density for ferromagnetic states e.g. (1,0,0) and (0,0,1) and the polar state (0,1,0) can be found in this way. This limit also reveals that mixed states do not exist at this limit as the sub-component density $n_0$ goes to zero (see Table 2 in \cite{Kanjilal_2020}). 
	Similarly, number density and energy density for $PM$ and $APM$ states given in  \cite{Kanjilal_2020} are ill-defined at $q=0$. So to find the number density and corresponding energy density for PM and APM states one can plug in $p=0$ and $q=0$ and solve Eq.(\ref{eq:rev7}-\ref{eq:rev8}) in the following manner.

    \subsubsection{Phase-matched state (PM)}
         As stated earlier for this stationary state, all the sub-components are populated and the relative phase, $\theta_r$ is zero. Using Thomas-Fermi approximation one can simplify Eq.(\ref{eq:rev7}) at $p,q=0$ (zero magnetic field) limit as,
        \begin{equation}\label{eq:rev10}
             U(\vec{r})+c_0n(\vec{r})-\mu+c_1\left(n_1+n_{-1}+2\sqrt{n_1n_{-1}}\right)=0.
        \end{equation}
        The same can be done for Eq.(\ref{eq:rev8}) after invoking the stationarity condition which yields two more equations describing the PM state, 
        \begin{equation}\label{eq:rev11}
            \begin{split}
                U(\vec{r})+c_0n(\vec{r})-\mu & + c_1(n_1-n_{-1})\\
                &+c_1n_0\left(1 +\sqrt{\dfrac{n_{-1}(\vec{r})}{n_{1}(\vec{r})}}\right)=0,
            \end{split}
        \end{equation}
     
        \begin{equation}\label{eq:rev12}
            \begin{split}
                U(\vec{r})+c_0n(\vec{r})-\mu & - c_1(n_1-n_{-1})\\
                &+c_1n_0\left(1 +\sqrt{\dfrac{n_{1}(\vec{r})}{n_{-1}(\vec{r})}}\right)=0.
            \end{split}
        \end{equation}
        So, we end up with three equations, Eqs.(\ref{eq:rev10}-\ref{eq:rev12}) to solve three unknown sub-component densities viz. $n_1$, $n_0$ and $n_{-1}$.
        The addition of Eq.(\ref{eq:rev11}) and Eq.(\ref{eq:rev12}) yields,
     
        \begin{equation}\label{eq:rev13}
            U(\vec{r})+c_0n(\vec{r})-\mu+ \dfrac{c_1n_0}{2\sqrt{n_1 n_{-1}}}\left(n_1+n_{-1}+2\sqrt{n_1n_{-1}}\right)=0.
        \end{equation}
        Note that,
        Eq.(\ref{eq:rev10}) and Eq.(\ref{eq:rev13}) describes the same PM state, as a result, a relation among the sub-component densities,
        \begin{equation}\label{eq:rev14}
            n_0=2\sqrt{n_1 n_{-1}},
        \end{equation}
        must hold true as all the sub-component densities are non-zero.
        Similarly subtracting Eq.(\ref{eq:rev11}) from Eq.(\ref{eq:rev12}) one gets to the same condition with no new information about the system.
     
        \par
     This simple relation among the sub-component densities, (Eq.(\ref{eq:rev14})) simplifies  Eq.(\ref{eq:rev10}) further to get an expression of total number density in terms of the external trapping potential,
        \begin{equation}\label{eq:rev15}
            n(r)=\dfrac{\mu-U(r)}{c_0+c_1}.
        \end{equation}
        The individual sub-component densities can be described as,
        \begin{equation}\label{eq:rev16}
            n_1=\dfrac{(n+\Tilde{m})^2}{4n},
        \end{equation}
        \begin{equation}\label{eq:rev17}
            n_0=\dfrac{n^2-\Tilde{m}^2}{2n},
        \end{equation}
        \begin{equation}\label{eq:rev18}
            n_{-1}=\dfrac{(n-\Tilde{m})^2}{4n},
        \end{equation}
        where, number density is defined as $n=n_1+n_0+n_{-1}$ and the magnetization density is defined as, $\Tilde{m}=n_1-n_{-1}$.
        \par
        The general expression for T-F approximated energy density can be written as, \cite{KAWAGUCHI2012253,computational}
        \begin{equation}\label{eq:rev19}
            \begin{split}
                e(\vec{r})=U(\vec{r})n(\vec{r})& +\dfrac{c_0}{2}n^2(\vec{r})+\dfrac{c_1}{2}\left(n_1-n_{-1}\right)^2\\
                &+ \dfrac{c_1}{2}n_0\left[n_1+n_{-1}+2\sqrt{n_1n_{-1}}cos\theta_r\right].
            \end{split}
        \end{equation}
        Sub-component density expressions, Eq.(\ref{eq:rev16}-\ref{eq:rev18}), allows one to get to the energy density for the phase-matched-state,
        \begin{equation}\label{eq:rev20}
            e_{PM}(r)=U(r)n(r)+\left(\dfrac{c_0}{2}+\dfrac{c_1}{4}\right)n^2(r)+\dfrac{c_1}{4}\Tilde{m}^2.
        \end{equation}
        As the sub-component densities and subsequently energy density depend on the magnetization density which is a free parameter in this analysis. To fix this, one may take resort to the minimization of energy density.
        \par
        It is interesting to note that, for anti-ferromagnetic type of spin-spin interaction, i.e. for $c_1>0$, $e_{PM}(\vec{r})$ is minimized at $\Tilde{m}=0$. Thus the sub-component densities become,
        $n_1=n/4$, $n_0=n/2$ and $n_{-1}=n/4$.
        \par
        Similarly, for ferromagnetic type of interaction at $c_1<0$, suggests that the energy density is minimized at $\Tilde{m}=n$. But, given the fact that all the sub-components are occupied (this is an essential condition for this stationary state, hence we can use all three Eqs.(\ref{eq:rev7}-\ref{eq:rev8})) the magnetization of the PM state can tend to the total number density but never be exactly equal to it i.e. $\Tilde{m}<n(\vec{r})$.  We will eventually observe that the ferromagnetic state dominate energetically in ferromagnetic interaction regime.

     \subsubsection{Anti-phase-matched state (APM)}
        The stationary state corresponding to $\theta_r=\pi$ and no empty sub-component is often referred to as the anti-phase-matched state or in short, the APM state. To find the number density and energy expressions we follow the same approach as described before.
        \par
        The stationary-phase conditions in Thomas-Fermi limit for the APM phase reduces Eqs.(\ref{eq:rev7}-\ref{eq:rev8}) into,
        \begin{equation}\label{eq:rev21}
     	    U(\vec{r})+c_0n(\vec{r})-\mu+c_1\left(n_1+n_{-1}-2\sqrt{n_1n_{-1}}\right)=0,
        \end{equation}
     
        \begin{equation}\label{eq:rev22}
     	    \begin{split}
     	        U(\vec{r})+c_0n(\vec{r})-\mu & + c_1(n_1-n_{-1})\\
     	        &+c_1n_0\left(1 -\sqrt{\dfrac{n_{-1}(\vec{r})}{n_{1}(\vec{r})}}\right)=0,
     	    \end{split}
        \end{equation}
     
        \begin{equation}\label{eq:rev23}
            \begin{split}
                U(\vec{r})+c_0n(\vec{r})-\mu & - c_1(n_1-n_{-1})\\
                &+c_1n_0\left(1 -\sqrt{\dfrac{n_{1}(\vec{r})}{n_{-1}(\vec{r})}}\right)=0.
            \end{split}
        \end{equation}
     
        By adding the last two equations we get,
     
        \begin{equation}\label{eq:rev24}
            U(\vec{r})+c_0n(\vec{r})-\mu -\dfrac{c_1n_0}{2\sqrt{n_1n_{-1}}}\left(n_1+n_{-1}-2\sqrt{n_1n_{-1}}\right)=0.
        \end{equation}
     
        So, the similarity check with Eq.(\ref{eq:rev21}) reveals that $n_0=-2\sqrt{n_1n_{-1}}$, which is a violation of assumption Eq.(\ref{eq:rev3}) because $\sqrt{n_m}$ is only positive by definition, while negativity can be absorbed in the phase factor $exp(-i\theta_m)$. So the only possible rescue is setting the whole term in the parentheses of Eq.(\ref{eq:rev24}) to zero. This gives the zero magnetization density condition, $\sqrt{n_1}=\sqrt{n_{-1}}$. So the anti-phase-matched state with non-zero magnetization is not possible in this limit. 
        \par
        Applying the constraint, $\sqrt{n_1}=\sqrt{n_{-1}}$, on Eq.(\ref{eq:rev21}) we find that the total number density varies independent of the spin-spin interaction coefficient i.e.
        \begin{equation}\label{eq:rev25}
            c_0n(\vec{r})=\mu-U(\vec{r})
        \end{equation}
        The other two equations, Eqs.(\ref{eq:rev22}-\ref{eq:rev23}) do not lead to anything new.
        \par
        By applying the constraint of zero magnetization density, energy density for the APM phase can be deduced from the general T-F approximated energy density expression, Eq.(\ref{eq:rev19}), as, 
        \begin{equation}\label{eq:rev26}
     	    e_{APM}(\vec{r})=U(\vec{r})\dfrac{(\mu-U(\vec{r}))}{c_0}+\dfrac{(\mu-U(\vec{r})^2)}{2c_0}
        \end{equation}
        Note that, in this case the sub-component number densities \emph{do not} come out from the stationarity conditions. Even the minimization of energy density does not fix the sub-component densities. So the APM state is a one parameter family of stationary states.
     
     
     \subsubsection{Ferromagnetic state}
        For the ferromagnetic phase, either $m_{F}=1$ or $m_{F}=-1$ is populated and the zeroth level is empty. The phase stationary equation corresponding to the populated level, which can be reduced from Eq.(\ref{eq:rev8}), gives the total density expression,
        \begin{equation}\label{eq:rev27}
            (c_0+c_1)n(\vec{r})=\mu-U(\vec{r}).
        \end{equation}
        By using the number density expression in the T-F approximated energy density (Eq.(\ref{eq:rev19})) we find,
        \begin{equation}\label{eq:rev28}
            e_{ferro}(\vec{r})=U(\vec{r})\dfrac{(\mu-U(\vec{r}))}{(c_0+c_1)}+\dfrac{(\mu-U(\vec{r}))^2}{2(c_0+c_1)}.
        \end{equation} 
        Note that, the number density expressions for ferromagnetic and PM state are identical, though the energy densities are different.
     
     \subsubsection{Polar and anti-ferromagnetic state}
         The stationary state corresponding to populated $m_{F}=0$ level and empty $m_F=1$, $m_F=-1$ levels is referred to as polar state.
        \par
        So, only the phase-stationary condition corresponding to the populated zeroth level is valid and Eq.(\ref{eq:rev7}) reduces to an expression of total number density, 
    
        \begin{equation}\label{eq:rev29}
            c_0n(\vec{r})=\mu-U(\vec{r}).
        \end{equation}
        The corresponding energy density becomes,
        \begin{equation}\label{eq:rev30}
            e_{polar}(\vec{r})=U(\vec{r})\dfrac{(\mu-U(\vec{r}))}{c_0}+\dfrac{(\mu-U(\vec{r}))^2}{2c_0},
        \end{equation}
         which is same as the expressions derived for the APM phase.
     
        \par
     
        For anti-ferromagnetic state only $m_{F}=0$ level is empty and the other two levels are populated.
        The number density and the energy density expressions are exactly same as the polar state,
        which is not so surprising given the fact that the quadratic Zeeman term is zero \cite{Kanjilal_2020}. The sub-component densities in this case are, $n_1=n/2$, $n_0=0$, $n_{-1}=n/2$, where $n$ is the total density.

     \section{T-F approximated total energy comparison at zero linear and quadratic Zeeman terms}
        Out of all the possible stationary states at zero magnetic field, the state having the lowest total energy would be selected as the ground state.
        \par
        For a fixed number of condensate particles, the chemical potential would be different for different stationary states. In the subsequent
        analysis we can assume an isotropic three-dimensional harmonic confinement for its simplicity.
     
     \subsection{T-F energy comparison for three dimensional isotropic confinement}
        Total number of particles $N$ can be found out by integrating the number density,
        \begin{equation}\label{eq:rev31}
     	    N=\int_{0}^{R}n(\vec{r})d\vec{r},
        \end{equation}
        where, $R$ is the T-F radius.
        \par
        As we have seen in the last section, the total number density for the ferromagnetic and $PM$ states vary in a similar fashion as shown in Eqs.(\ref{eq:rev15}, \ref{eq:rev27}). So for these states,
        \begin{equation}\label{eq:rev32}
            N=\int_{0}^{R_1}4\pi r^2\dfrac{\mu_1-M\omega^2r^2/2}{(c_0+c_1)}dr
        \end{equation}
        \begin{equation}\label{eq:rev33}
            \implies N=\dfrac{4\pi M\omega^2 R_1^5}{15(c_0+c_1)}
        \end{equation}
        where, the chemical potential is related to the T-F radius $R_1$ as, $\mu_1=M\omega^2R_1^2/2$ \cite{2003bose} and $M$ is the mass of the condensate particle and $\omega$ is the trapping frequency.
        \par
     
        So, the chemical potential for the PM and ferromagnetic states assume the form,
        \begin{equation}\label{eq:rev34}
     	    \mu_1=\dfrac{M\omega^2}{2}\left[\dfrac{15(c_0+c_1)N}{4\pi M\omega^2}\right]^{\dfrac{2}{5}}.
        \end{equation}
	    Similarly for polar, anti-ferromagnetic and APM states the total number density varies differently, as Eq.(\ref{eq:rev25}). The chemical potential for these states can be expressed as,
	    \begin{equation}\label{eq:rev35}
	        \mu_2=\dfrac{M\omega^2}{2}\left[\dfrac{15c_0N}{4\pi M\omega^2}\right]^{\dfrac{2}{5}}.
	    \end{equation}
	
	\subsubsection{Comparison of total energy for ferromagnetic type of spin-spin interaction}
	    Total energy for all the stationary states can be found out by integrating the corresponding energy densities.
	    The total energy for the ferromagnetic state can be expressed as,
	    \begin{equation}\label{eq:rev36}
		    E_{ferro}=\dfrac{2\pi M^2\omega^4}{21}\left[\dfrac{15N}{4\pi M\omega^2}\right]^{7/5}(c_0+c_1)^{2/5},
	    \end{equation}
	    where, we have made use of Eqs.(\ref{eq:rev33}-\ref{eq:rev34}) to express the energy only in terms of the total number of condensate particles $N$.
	    \par
	    The total energy for the PM state is,
	    \begin{equation}\label{eq:rev37}
	        \begin{split}
	        E_{PM}=&\int_{0}^{R_{PM}}U(r)n(r)4\pi r^2 dr+\int_{0}^{R_{PM}}\dfrac{c_1}{4}\Tilde{m}^2 4\pi r^2 dr\\
	        &+\int_{0}^{R_{PM}}\left(\dfrac{c_0}{2}+\dfrac{c_1}{4}\right)n^2(r)4\pi r^2 dr.     
	        \end{split}
	    \end{equation}
	    Now this can be written as,
	    \begin{equation}\label{eq:rev38}
	        \begin{split}
	        E_{PM}=\int_{0}^{R_{PM}}&U(r)n(r)4\pi r^2 dr+\\
	        &\int_{0}^{R_{PM}}\left(\dfrac{c_0}{2}+\dfrac{c_1}{2}\right)n^2(r)4\pi r^2 dr\\
	        &\quad+(\mathcal{I}-1)\int_{0}^{R_{PM}}\dfrac{c_1}{4}n^2 4\pi r^2 dr
	        \end{split}
	    \end{equation}
	    where,
	    \begin{equation}\label{eq:rev39}
	        \mathcal{I}=\dfrac{\int_{0}^{R_{PM}}\Tilde{m}^2 4\pi r^2 dr}{\int_{0}^{R_{PM}}n^2 4\pi r^2 dr}
	    \end{equation}
	    
	    As a result the $PM$ state energy can be written as,
	    \begin{equation}\label{eq:rev40}
	        \begin{split}
	        E_{PM}=E_{ferro}+(\mathcal{I}-1)\int_{0}^{R_{PM}}\dfrac{c_1}{4}n^2 4\pi r^2 dr
	        \end{split}
	    \end{equation}
	
	    \par
	    Now with an usual configuration, $1>\mathcal{I}>0$. So in this limit, as the second term is going to increase the total energy at $c_1<0$ on top of the ferromagnetic energy. So ferromagnetic state will be the lower energy state.
	    Similarly, the total energy for polar, anti-ferromagnetic and APM state is expressed as,
	    \begin{equation}\label{eq:rev41}
	        E_{APM}=\dfrac{2\pi M^2\omega^4}{21}\left[\dfrac{15N}{4\pi M\omega^2}\right]^{7/5}(c_0)^{2/5}.
	    \end{equation} 
	    Note that, as discussed earlier the number density, energy density and chemical potential, all these quantities are similar for APM, polar and anti-ferromagnetic state. So it is not surprising that the total energy would also be same for these three states.

	    For $c_1<0$, we have already discussed that the phase-matched state does not win energetically over the ferromagnetic state. So out of two possible choices, Eq.(\ref{eq:rev36}) and Eq.(\ref{eq:rev41}) suggests that $c_1$ being negative would only decrease the ferromagnetic energy thus making it the ground state in this limit.


	\subsubsection{Comparison of total energy for anti-ferromagnetic type of spin-spin interaction}
	    When the spin-spin interaction is positive, the total energy for the phase-matched state can be calculated to be,
	    \begin{equation}\label{eq:rev42}
	        E_{PM}=E_{ferro}-\dfrac{2\pi M^2\omega^4}{105}\left[\dfrac{15N}{4\pi M\omega^2}\right]^{7/5}c_1(c_0+c_1)^{-3/5}
	    \end{equation}
	
	    For $c_1>0$, comparison of the total energy of the ferromagnetic state (Eq.(\ref{eq:rev36})), anti-phase-matched state (Eq.(\ref{eq:rev41})) and the PM state reveals that, $E_{ferro}>E_{APM}$ and $E_{ferro}>E_{PM}$. So it is obvious that the ferromagnetic state can not be the ground state in this limit. To further explore the competition between the $APM$/polar/AF and $PM$ states
        we choose realistic values of $c_1$ and $c_0$ for $^{23}Na$. Let the total number of the particle be $10^5$, because it is well known that for large condensates T-F prediction gives fairly accurate results. For this case,
        \begin{enumerate}[\indent {}]
            \item $c_1=2.415\times10^{-19}\hspace{0.25em} Hz\hspace{0.25em} m^3,$
            \item $c_0=149.89\times10^{-19}\hspace{0.25em} Hz\hspace{0.25em} m^3,$
            \item $\omega=2\pi\times100 \hspace{0.25em} Hz,$
 	        \item $\dfrac{1}{2} m \omega^2=1.13725464\times10^{13} \hspace{0.25em} Hz\hspace{0.25em} m^{-2}.$
 	    \end{enumerate}
        Thus, chemical potential $\mu_2=\dfrac{M\omega^2}{2}\left[\dfrac{15c_0N}{4\pi M \omega^2}\right]^{2/5}=1033.15$ $Hz$, which roughly is $50$ $nK$. This corresponds to the APM, polar and anti-ferromagnetic state. For the PM and ferromagnetic state the chemical potential is $\mu_1=\dfrac{M\omega^2}{2}\left[\dfrac{15(c_0+c_1)N}{4\pi M \omega^2}\right]^{2/5}=1039.04$ $Hz$.
        \par 
    
        The total energy for all possible stationary states can be calculated by using the parameter values shown earlier, 
        \begin{enumerate}[\indent {}]
            \item $E_{APM} \simeq 7.38\times10^7\hspace{0.25em} Hz$,
            \item $E_{PM}\simeq 7.40\times10^7\hspace{0.25em} Hz$,
            \item $E_{ferro} \simeq 7.42\times10^7\hspace{0.25em} Hz$.
        \end{enumerate}
        So, we see that for a condensate at zero magnetic field under a 3-D harmonic confinement, the APM state is the lowest energy stationary state when the spin-spin interaction is of anti-ferromagnetic type. Note that, not only the APM state but also polar and anti-ferromagnetic states are equally likely to be the ground state in this limit. However, the energy difference of the PM state and the APM state is just $0.3$\% of that of the PM state. This is an extremely small energy difference and the non-consideration of kinetic energy could have significant consequences especially for a system of smaller number of particles where T-F approximation is known not to be a good one.
        \par
        This warrants a closer look at the situation with beyond T-F approximation.
        
        \subsection{T-F energy comparison in a quasi-one dimensional harmonic confinement}
        The same energy comparison can be done in a quasi-one dimensional harmonic trap for which the transverse trapping frequency (say, along $y$- and $z$-axis) $\sqrt{\omega_y\omega_z}>\omega_x$ is much greater than the longitudinal trapping frequency along $x$-axis. For calculation purposes lets assume the oscillator length along the transverse direction, $l_{yz}=\sqrt{\hbar/M\omega_{yz}}$ is 0.59 $\micro m$ and the oscillator length along the $x$-direction $l_x=\sqrt{\hbar/M\omega_{x}}$
        is 2.965 $\micro m$. 
        \par
        \underline{\textit{Ferromagnetic type of interaction}} The analysis given for three dimensional isotropic confinement holds for the quasi-one dimensional condensate as well.
        \par
        For 1-D condensate the total energy of the APM or polar or anti-ferromagnetic stationary state can be written as,
        \begin{equation}\label{eq:rev43}
            E^{1D}_{APM}=\dfrac{\pi l^2_{yz}}{5}M^2\omega_x^4\left[\dfrac{3N}{2\pi l^2_{yz} M\omega_x^2}\right]^{5/3}(c_0)^{2/3}.
        \end{equation}
        Similarly the total energy for ferromagnetic state is,
        \begin{equation}\label{eq:rev44}
            E^{1D}_{ferro}=\dfrac{\pi l^2_{yz}}{5}M^2\omega_x^4\left[\dfrac{3N}{2\pi l^2_{yz} M\omega_x^2}\right]^{5/3}(c_0+c_1)^{2/3}.
        \end{equation}
        The total energy for PM state for ferromagnetic type of interaction can be written in a similar way as Eq.(\ref{eq:rev40}),
        \begin{equation}\label{eq:rev45}
            \begin{split}
	            E_{PM}^{1D}=E_{ferro}^{1D}+(\mathcal{I}^{1D}-1)2\pi l_{yz}^2\int_{0}^{X_{PM}}\dfrac{c_1}{4}n^2(x) dx
	        \end{split}
        \end{equation}
        where, $X_{PM}$ is the T-F radius in one dimensional harmonic confinement for the PM state. The term $\mathcal{I}^{1D}$ in this 1-D case is defined as,
        \begin{equation}\label{eq:rev46}
	        \mathcal{I}^{1D}=\dfrac{\int_{0}^{X_{PM}}\Tilde{m}^2  dx}{\int_{0}^{X_{PM}}n^2(x) dx}.
	    \end{equation}
        It is evident that $0<\mathcal{I}^{1D}<1$. For ferromagnetic type of interaction ($c_1<0$), $E_{PM}>E_{ferro}$ and $E_{APM}>E_{ferro}$. So clearly in 1-D condensate with ferromagnetic type of interaction the T-F approximation suggests that the ferromagnetic state is the lowest energy stationary state.
        \par
        \underline{\textit{Anti-ferromagnetic type of interaction:}} 
        For anti ferromagnetic type of interaction the parameter $\Tilde{m}=0$
        for the PM state as a result $\mathcal{I}^{1D}=0$. So the total energy for the PM state can be calculated as,
        \begin{equation}\label{eq:rev47}
            E_{PM}^{1D}=E_{ferro}^{1D}-\dfrac{\pi l_{yz}^2 }{15}M^2\omega_x^4\left[\dfrac{3N}{2\pi l_{yz}^2 M\omega_x^2}\right]^{5/3}\dfrac{c_1}{(c_0+c_1)^{1/3}}
        \end{equation}
        So, using the specific choice of numerical values of all the parameters we find, 
        \begin{enumerate}[\indent {}]
            \item $ E_{APM}^{1D}\simeq8.648\times10^8\hspace{0.25em} Hz,$
            \item $E_{PM}^{1D}\simeq8.695\times10^8 \hspace{0.25em} Hz,$
            \item $E_{ferro}^{1D}\simeq8.74\times10^8 \hspace{0.25em} Hz,$
        \end{enumerate}
        for condensate having a total of $N=10^5$ number of particles. So for an 1-D condensate with anti ferromagnetic interaction the APM, polar and the anti-ferromagnetic state are equally likely to be the ground state but the energy difference with the PM state becomes approximately $0.5\%$ which is also fairly small. This T-F energy difference is independent of $N$. So for small condensate where T-F does not produce fairly accurate result, it is natural to suspect that the consideration of the full profile of the condensate might affect this conclusion.
    
    \section{Variational approach: anti-ferromagnetic type of interaction}
    
    Previously we saw that the T-F limit predicts that, for an anti-ferromagnetic condensate under 3-D isotropic harmonic trapping, the anti-phase-matched or polar or the anti-ferromagnetic state becomes the ground state when $p=0$ and $q=0$. But, the energy difference between the phase-matched-state and the three degenerate states is only about $0.3\%$. The full profile of the condensate may tell a different story than what the T-F approximation provides. Thus, there is a need to consider the full profile of the condensate in place of the T-F approximated profile. \par

    Firstly, we restate the GP equations for the stationarity of phases at $p=0$ and $q=0$,
    \begin{equation}\label{eq:rev48}
        \lbrace \mathcal{H}-\mu+c_1\left(n_1+n_{-1}+2 \sqrt{n_{-1}n_1}\cos\theta_r\right)\rbrace \sqrt{n_0(\vec{r})}=0,
    \end{equation}
    \begin{equation}\label{eq:rev49}
        \begin{split}
            \lbrace \mathcal{H}-\mu & \pm c_1\left(n_1-n_{-1}\right)+q \mp p\rbrace\sqrt{n_{\pm1}(\vec{r})}\\
            &+c_1n_0\left(\sqrt{n_{\pm1}(\vec{r})} +\sqrt{n_{\mp1}(\vec{r})}\cos\theta_r\right)=0.
        \end{split}
    \end{equation}
    The interaction parameter and the number densities can be rescaled as,
    \begin{equation}\label{eq:rev50}
  	    c_0=(4/3) \pi l_0^3\lambda_0\hbar\omega, \quad c_1=(4/3) \pi l_0^3\lambda_1\hbar\omega,
    \end{equation}
    \begin{equation}\label{eq:rev51}
        u_m=(4/3) \pi l_0^3 \lambda_0 n_m, \quad r=l_0\eta
    \end{equation}
    where, $l_0=\sqrt{\hbar/(m\omega)}$ is the oscillator length scale. As a result of this transformation, the parameters $\lambda_0$, $\lambda_1$, $\eta$ and $u_m$ become all dimensionless.
    \par
	Thus, Eq.(\ref{eq:rev48}) and Eq.(\ref{eq:rev49}) can be rewritten in dimensionless form as,
    \begin{equation}\label{eq:rev52}
	    \begin{split}
	        \bigg\{ -\dfrac{1}{2}\dfrac{1}{\eta^2}&\dfrac{d}{d\eta}(\eta^2\dfrac{d}{d\eta})+\dfrac{1}{2}\eta^2+ u-\mu'\\
	        &+\lambda'_1\left(u_1+u_{-1}+2 \sqrt{u_{-1}u_1}\cos\theta_r\right) \bigg\} \sqrt{u_0}=0,
	    \end{split}
	\end{equation} 
	\begin{equation}\label{eq:rev53}
		\begin{split}
		\bigg\{ -\dfrac{1}{2}\dfrac{1}{\eta^2}\dfrac{d}{d\eta}(\eta^2\dfrac{d}{d\eta})+&\dfrac{1}{2}\eta^2+ u-\mu' \pm \lambda'_1\left(u_1-u_{-1}\right)\bigg\}\sqrt{u_{\pm1}}\\
		&+\lambda'_1u_0\left(\sqrt{u_{\pm1}} +\sqrt{u_{\mp1}}\cos\theta_r\right)=0.
		\end{split}
	\end{equation}
    
	where, $\lambda_1'=\lambda_1/\lambda_0$ and $\mu'=\mu/(\hbar \omega)$. The total number density $u$ is the sum of all the sub-component density, i.e. $u=u_1+u_0+u_{-1}$. 
		
	\subsection{Variational method}
	As we are dealing with an 3-D isotropic harmonic trap, one may get the number density at the high density central region by incorporating the fact that the kinetic energy, compared to interaction energy will have negligible contribution. So, the first terms of both Eqs.(\ref{eq:rev52}-\ref{eq:rev53}) can be neglected and the corresponding total number density in the high density region can be found out for all the stationary states as a function of $\mu'$ and $\eta$.
	Similarly, at low density region the kinetic energy will have greater contribution as compared to the interaction energy and we assume the number density to be gaussian-like in this region in analogy with the ground state under harmonic oscillator potential.  
	\par
	We assume the high density (central region) and the low density (outer region) number density solution to match smoothly at $\eta=\eta_0$. To summarize,
 	\begin{enumerate}[\indent {}]
 		\item{$u^{in}(\eta)=f(\mu',\eta)$ for $\eta<\eta_0$},
 		\item{$u^{out}(\eta)=(a+c \eta +d \eta^2) \exp\left(-\dfrac{\eta^2}{b}\right)$ for $\eta\geq\eta_0$};
 	\end{enumerate}
 	given, $\sqrt{u}$, and its first, second and third derivatives on both sides should be equal at $\eta=\eta_0$. As discussed, the exact functional form $f(\mu',\eta)$ is different for different stationary states and can be found from the solution of the Eqs.(\ref{eq:rev52}-\ref{eq:rev53}) by neglecting the kinetic energy term. Following the smooth-matching-condition, $a$, $b$, $c$ and $d$ can be written in terms of $\eta_0$ and $\mu'$. The parameter $\mu'$ can be fixed from the total number conservation as a function of number of condensate particles N and matching point $\eta_0$. As a result, for a given number of condensate particles, the total energy can be calculated easily. Note that, the total energy then becomes only a function of $\eta_0$ and minimization of the total energy fixes the free parameter $\eta_0$.
 	\par
 	Note that, it is necessary to consider the smoothness up to the third derivative so that the kinetic energy on both sides of $\eta_0$ matches smoothly hence, we get a smooth energy density profile as well. This procedure takes into account of all the sub-component densities and in general, the coefficients $a$, $b$, $c$ and $d$ may be different for different sub-components. This signifies the multi-modal nature of this variational approach and the method is applicable even if the sub-components do not follow a single spatial profile. 
 	\par
 	Following this procedure for polar (or AF or APM) and PM state would allow us to compare the total energy of these two states which is our main purpose.

 	\subsubsection{Polar state}
 	    For polar state, only the zeroth sub-component is populated and the others are empty so $u=u_0$ and $u_{\pm1}=0$. Neglecting the kinetic energy term in Eqs.(\ref{eq:rev52}-\ref{eq:rev53}) the high density solution is found out as,
 	    \begin{equation}\label{eq:rev54}
 	        u_{pol}^{in}(\eta)=\mu'-\eta^2/2
 	    \end{equation}
 	    The number density matching condition, i.e., $\sqrt{u_{pol}^{in}(\eta_0)}=\sqrt{u_{pol}^{out}(\eta_0)}$ gives the estimate of $a$ as function of $b$, $c$, $d$ and $\eta_0$ as,
 	    \begin{equation}\label{eq:rev55}
 		    a=\left(\mu'-\eta_0^2/2\right)exp\left(\dfrac{\eta_0^2}{b}\right)-c \eta_0 -d \eta_0^2.
 	    \end{equation}
 	    Matching the slope of $\sqrt{u_{pol}(\eta)}$ in outer and inner region at $\eta_0$ gives $d$ as a function of $c$, $b$ and $\eta_0$,
 	    \begin{equation}\label{eq:rev56}
 		    d=exp\left(\dfrac{\eta_0^2}{b}\right)\left(\dfrac{\mu'-\eta_0^2/2}{b}-\dfrac{1}{2}\right)-\dfrac{c}{2\eta_0}.
 	    \end{equation}
 	    The second derivative matching condition at $\eta_0$ gives $c$ in terms of $b$ and $\eta_0$,
 	    \begin{equation}\label{eq:rev57}
 	        c=\dfrac{2exp\left(\dfrac{\eta_0^2}{b}\right)\eta_0^3(2b-2\mu'+\eta_0^2)}{b^2}.
 	    \end{equation}
 	    The parameter $b$ is found out from the third derivative matching condition,
 	    \begin{equation}\label{eq:rev58a}
 	        b=\dfrac{1}{12}\left(6 \mu'-9\eta_0^2+\sqrt{36\mu^2-12 \mu'\eta_0^2+33\eta_0^4}\right),
 	    \end{equation}
 	   where, the parameter $b$ is written only in terms of the parameter $\mu'$ and $\eta_0$. Plugging in the expression of $b$ in Eq.(\ref{eq:rev57}), the parameter $c$ can also be expressed in terms of the parameters $\mu'$ and $\eta_0$. Similarly, $a$ and $d$ can be expressed in the same way. Therefore, the number density in the outer region can be written only in terms of the parameters $\mu'$ and $\eta_0$.
 	    To determine $\mu'$, we use the total number conservation condition,
 	    \begin{equation}\label{eq:rev58}
 	        \int^{\eta_0}_0 u^{in}_{pol}\eta^2 d\eta+\int^{\infty}_{\eta_0} u^{out}_{pol}\eta^2 d\eta =\lambda_0 N/3.
 	    \end{equation}
        Following the integration and simplifying further we get the equation,
        \begin{equation}\label{eq:rev59}
            \begin{split}
             &\dfrac{1}{192 k^{5/2}} \Bigg[12 \sqrt{k}
            \eta_0 \bigg(168 k \mu'^3 + 4 \mu'^2 (-113 k + 336 \mu') \eta_0^2\\
            &+ 18 (23 k - 176 \mu') \mu' \eta_0^4 + (53 k + 816 \mu') \eta_0^6+ 216 \eta_0^8\bigg)\\
            & +6 \exp{\dfrac{12 \eta_0^2}{k}} \sqrt{
            3 \pi} \bigg(336 k \mu'^4 + 128 \mu'^3 (-10 k + 21 \mu') \eta_0^2\\
            &+8 (269 k - 1168 \mu') \mu'^2 \eta_0^4 + 16 \mu' (-115 k + 872 \mu') \eta_0^6 \\
            &+ (473 k - 7904 \mu') \eta_0^8 + 1464 \eta_0^{10}\bigg)
            + \exp{\dfrac{12 \eta_0^2}{k}} k \sqrt{ 3 \pi}\\ 
            &\bigg(-168 k \mu'^3 + 4 (125 k - 336 \mu') \mu'^2 \eta_0^2\\
            &+ 2 \mu' (-311 k + 1776 \mu') \eta_0^4 + 3 (61 k - 784 \mu') \eta_0^6\\
            &+456 \eta_0^8\bigg) Erf\Big(\dfrac{2 \sqrt{3} \eta_0}{\sqrt{k}}\Big)\bigg)\Bigg]
            + \dfrac{\mu' \eta_0^3}{3} - \dfrac{\eta_0^5}{10}=\lambda_0N/3,
            \end{split}
        \end{equation}
    where, $k=6 \mu'-9\eta_0^2+\sqrt{36\mu^2-12 \mu'\eta_0^2+33\eta_0^4}$ and $Erf\Big(\dfrac{2 \sqrt{3} \eta_0}{\sqrt{k}}\Big)$ is the error function and $\lambda_0$ and $N$ are known for a condensate. Thus, Eq.(\ref{eq:rev59}) allows us to estimate $\mu'$ numerically for different values of $\eta_0$ and $N$.
    \par
    Now for a particular anti-ferromagnetic condensate (for a given $N$) one can calculate the total energy for different values $\eta_0$ once we find the respective $\mu'(\eta_0)$ from Eq.(\ref{eq:rev59}). The polar-state energy density can be written in these dimensionless parameters as (using Eqs.(\ref{eq:rev50}-\ref{eq:rev51}) in  Eq.(\ref{eq:rev9})),
    
    \begin{equation}\label{eq:rev60}
        \begin{split}
            e_{pol}(u(\eta))=\dfrac{3\hbar\omega }{2\lambda_0}\Big[-\sqrt{u(\eta)}\dfrac{1}{\eta^2}&\dfrac{d}{d\eta}(\eta^2\dfrac{d}{d\eta}\sqrt{u(\eta)})\\
            &+\eta^2u(\eta)+u^2(\eta)\Big].
        \end{split}
    \end{equation} 
    
    The total energy for polar state can be found out by integrating the energy density,
    \begin{equation}\label{eq:rev61}
		E_{pol}(\eta_0)=\int_{0}^{\eta_0}d\eta  \eta^2 e_{pol}(u^{in}(\eta))+\int_{\eta_0}^{\infty}d\eta \eta^2 e_{pol}(u^{out}(\eta)).
    \end{equation}
   
    \par
   
    Now the total energy of the polar state can be calculated for different values of the matching point and $\dfrac{dE_{pol}(\eta_0)}{d\eta_0}=0$ fixes the value of $\eta_0$ as well as the minimum possible total energy for the polar state.
	\par
	\begin{figure}[ht]
		\subfloat[$N=10^3$\label{subfig-1:Polar state n3}]{%
			\includegraphics[width=0.23\textwidth]{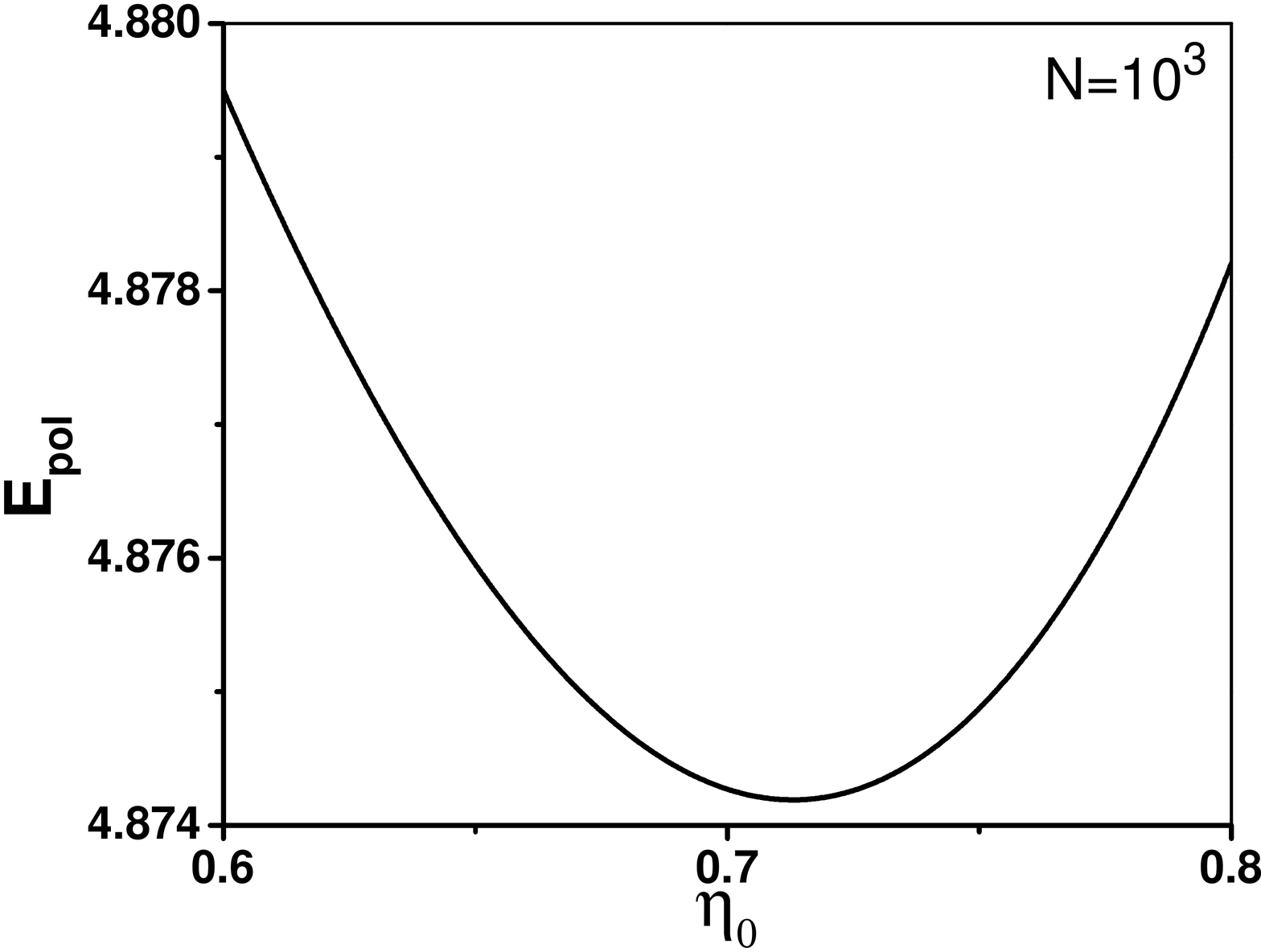}
		}
		\subfloat[$N=10^4$\label{subfig-1:Polar state n4}]{%
			\includegraphics[width=0.23\textwidth]{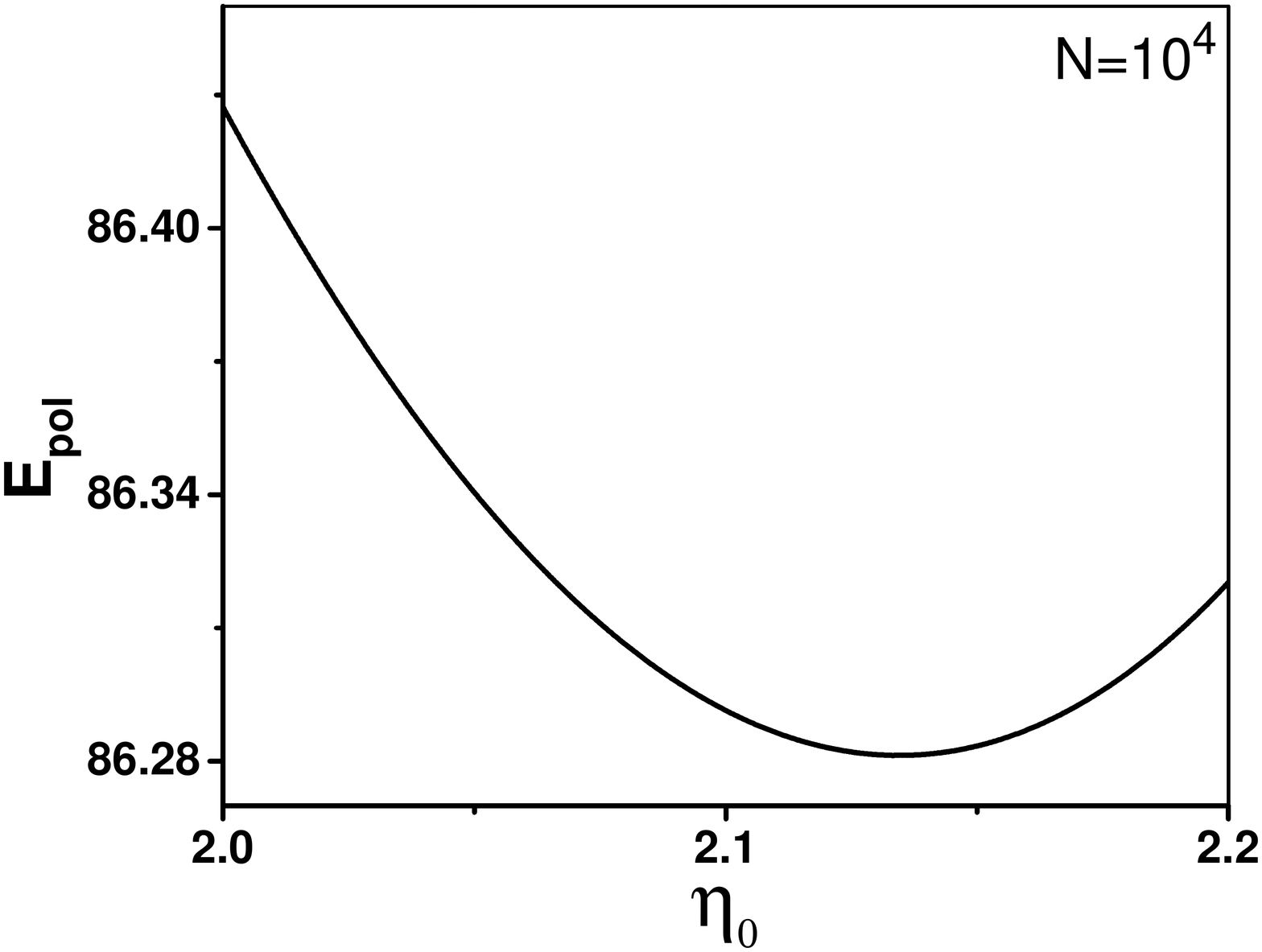}
		}	
		\caption{Total energy of polar state is calculated for a range of $\eta_0$ (the position at which the number densities at high and low density region smoothly connects). The free parameter $\eta_0$ is fixed at the value corresponding to the minimum of energy. The actual total energy in 3-D can be found out by multiplying the factor $3\hbar \omega/(2\lambda_0)$ with the values shown in $y-axis$. For the purpose of visualization this scaling factor for energy is used in all subsequent figures (of the 3-D case). Subfig-(a) shows the existence of minima of the total energy of the polar state with respect to the matching point $\eta_0$ when the total number of condensate particles is 1,000 and the T-F radius is $\simeq$1.81. Similarly for 10,000 number of condensate particles the energy minima condition in subfig-(b) determines the matching point $\eta_0$. Its no surprise that the $\eta_0$ shifts more towards to the T-F radius, which in this case is $\simeq$2.283. As the particle number is increased T-F approximation produces  fairly accurate results.}
		\label{fig:polar}
	\end{figure}
	\par
    It is no surprise that the variational method (VM), by introduction of Gaussian tail part in the low density region, does not lift the degeneracy of the polar, anti-ferromagnetic and APM phase as both $p=0$ and $q=0$. 
	
	\subsubsection{Phase-matched state}
	     
	     Following a similar approach, first we get the sub-component densities for PM state in the central region where the kinetic energy contribution can be neglected. Note that all the sub-components are populated for PM state and the solution of Eqs.(\ref{eq:rev52}-\ref{eq:rev53}) allows to write the sub-component densities in terms of the total number density of the PM state,
	    \begin{equation}\label{eq:rev62}
	        u_{\pm1}^{in}= u^{in}/4, \quad u_0^{in}=u^{in}/2,
	    \end{equation}
	    where, $u^{in}_{m}$ for $m=1,0,-1$ represents the sub-component densities in high density inner region and $u^{in}$ is the total density. The total density in this region ($\eta<\eta_0$) comes out to be,
	    \begin{equation}\label{eq:rev63}
	        u^{in}= \dfrac{\mu'-\eta^2/2}{(1+\lambda_1')}.    
	    \end{equation}
	     The sub-component densities in the outer region where the kinetic energy contribution is significant is taken as,
	    \begin{equation}\label{eq:rev64}
	        u_m^{out}(\eta) = (a_m+c_m\eta+d_m\eta^2) \exp{(-\eta^2/b_m)} \hspace{0.25em} for \hspace{0.25em} \eta\geq\eta_0.
	    \end{equation}
         \begin{figure}[htbp]
	    	\subfloat[$N=10^3$\label{subfig-1:PM state n3}]{%
	    		\includegraphics[width=0.23\textwidth]{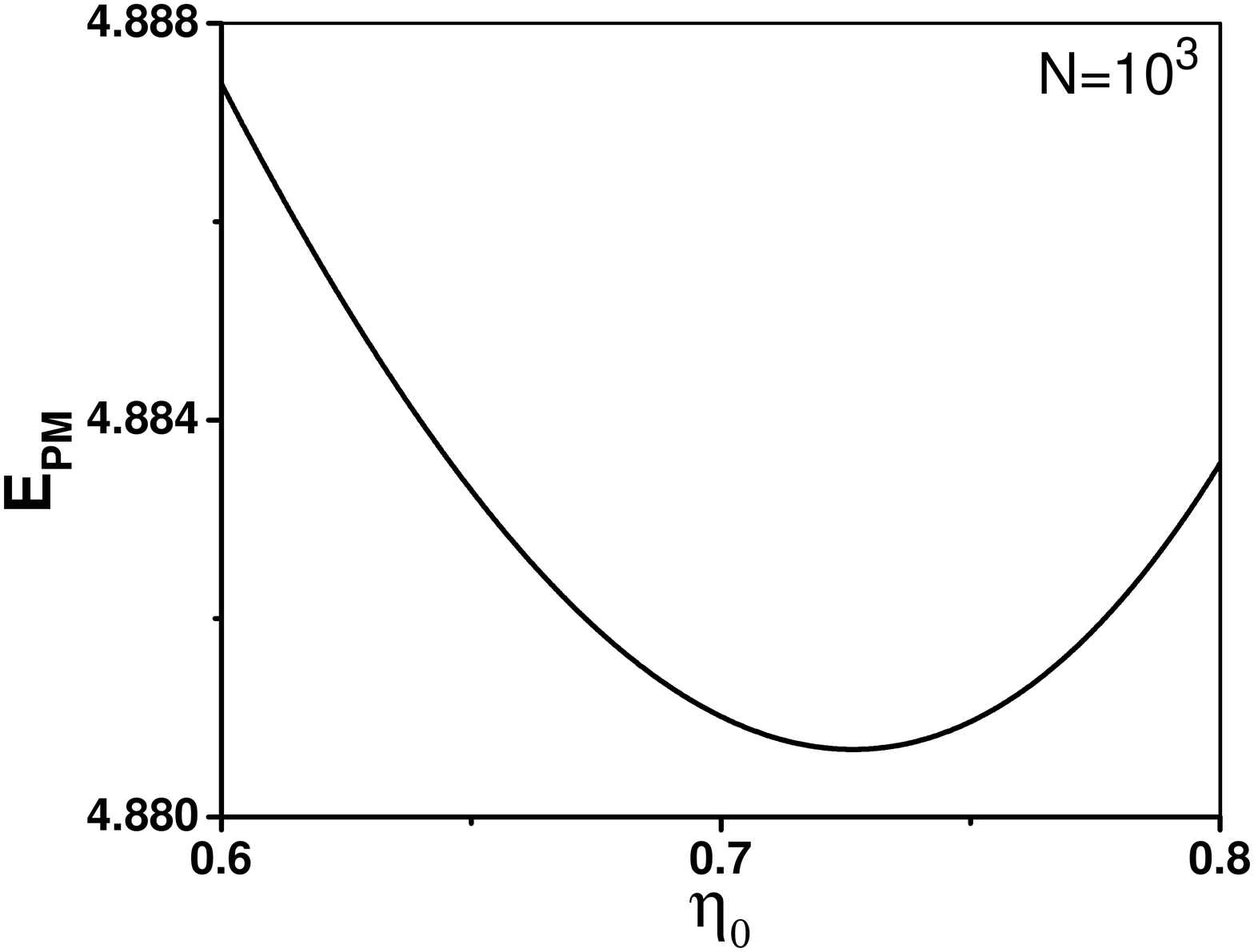}
	    	}
	    	\subfloat[$N=10^4$\label{subfig-1:PM state n4}]{%
	    		\includegraphics[width=0.23\textwidth]{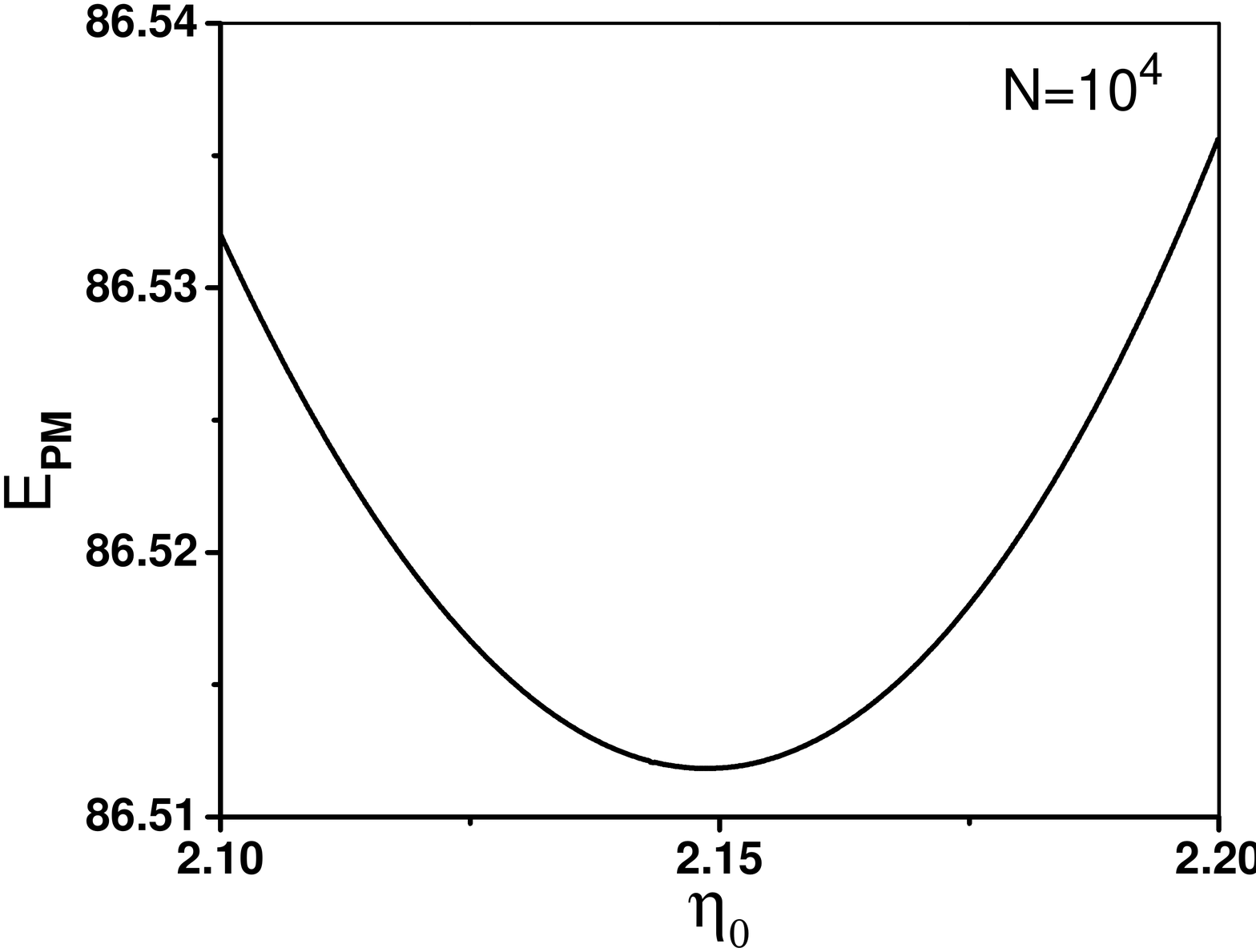}
	    	}	
	    	\caption{Total energy of the $PM$-state is calculated for varying $\eta_0$. The minimization of total energy fixes the free parameter $\eta_0$, for example, for 1,000 condensate particles (subfig-(a)) we find the $\eta_0$ is fixed at 0.727 (approx.) corresponding to the minimum value of total energy (4.881) while the T-F radius is approximately at 1.815. Similarly for 10,000 condensate particles (subfig-(b)) we find $\eta_0$ is fixed at 2.149 (approx.) corresponding to the minimization of energy at 86.512 while the T-F radius is 2.877 (approx.).}
	    	\label{fig:PM}
	    \end{figure}
	    
	    Following the same procedure of employing smooth matching condition, the coefficients $a_m$, $b_m$, $c_m$ and $d_m$ can be found out for the PM state. 
	    \begin{equation}\label{eq:rev65}
	        b_0=b_{\pm1}=b, 
	    \end{equation}
	    \begin{equation}\label{eq:rev66}
	         c_{\pm1}=\dfrac{c_{0}}{2}=\dfrac{c}{4(1+\lambda_1')}, 
	    \end{equation}
	    \begin{equation}
 		    d_{\pm1}=\dfrac{d_0}{2}=\dfrac{d}{4(1+\lambda_1')},
 	    \end{equation}
 	    \begin{equation}
 		    a_{\pm1}=\dfrac{a_0}{2}=\dfrac{a}{4(1+\lambda_1')},
 	    \end{equation}
 	    where, the parameters $a$, $b$, $c$, $d$ are found out to be of similar expression as Eqs.(\ref{eq:rev55}-\ref{eq:rev58a}) with $\mu'$ being different than that of polar state.
	    The parameter $\mu'$ for the PM state can be found out from the total number conservation,
	    \begin{equation}\label{eq:rev67}
 	        \int^{\eta_0}_0 u_{PM}^{in} \eta^2 d\eta+\int^{\infty}_{\eta_0} u_{PM}^{out} \eta^2 d\eta =\lambda_0 N/3,
 	    \end{equation}
 	    where, $u^{out}_{PM}=u_1^{out}(\eta)+u_0^{out}(\eta)+u_{-1}^{out}(\eta)$ is the total number density at $\eta\geq\eta_0$.
	    Following the integration one arrives at,
        \begin{equation}\label{eq:rev68}
            \begin{split}
                 &\dfrac{1}{192 k^{5/2}} \Bigg[12 \sqrt{k}
            \eta_0 \bigg(168 k \mu'^3 + 4 \mu'^2 (-113 k + 336 \mu') \eta_0^2\\
            &+ 18 (23 k - 176 \mu') \mu' \eta_0^4 + (53 k + 816 \mu') \eta_0^6+ 216 \eta_0^8\bigg)\\
            & +6 \exp{\dfrac{12 \eta_0^2}{k}} \sqrt{
            3 \pi} \bigg(336 k \mu'^4 + 128 \mu'^3 (-10 k + 21 \mu') \eta_0^2\\
            &+8 (269 k - 1168 \mu') \mu'^2 \eta_0^4 + 16 \mu' (-115 k + 872 \mu') \eta_0^6 \\
            &+ (473 k - 7904 \mu') \eta_0^8 + 1464 \eta_0^{10}\bigg)
            + \exp{\dfrac{12 \eta_0^2}{k}} k \sqrt{ 3 \pi}\\ 
            &\bigg(-168 k \mu'^3 + 4 (125 k - 336 \mu') \mu'^2 \eta_0^2\\
            &+ 2 \mu' (-311 k + 1776 \mu') \eta_0^4 + 3 (61 k - 784 \mu') \eta_0^6\\
            &+456 \eta_0^8\bigg) Erf\Big(\dfrac{2 \sqrt{3} \eta_0}{\sqrt{k}}\Big)\bigg)\Bigg]
            + \dfrac{\mu' \eta_0^3}{3} - \dfrac{\eta_0^5}{10}\\
            &\qquad\qquad\qquad\qquad\qquad\qquad =(1+\lambda_1')\lambda_0N/3,
            \end{split}
        \end{equation}
        where, $k=6 \mu'-9\eta_0^2+\sqrt{36\mu^2-12 \mu'\eta_0^2+33\eta_0^4}$. This equation is similar to Eq.(\ref{eq:rev59}) except for the coefficient $(1+\lambda_1')$ on the r.h.s. The parameter $\mu'$ for the PM state can be computed from this equation for different values of $N$ and $\eta_0$.
	
	    The energy density for PM state for 3-D harmonic confinement is,
        \begin{equation}\label{eq:rev69}
            \begin{split}
                e_{PM}=\dfrac{3\hbar\omega_x}{2\lambda_0}&\Big[-\sqrt{u(\eta)}\dfrac{1}{\eta^2}\dfrac{d}{d\eta}(\eta^2\dfrac{d}{d\eta}\sqrt{u(\eta)})\\
                &+\eta^2u(\eta)+(1+\lambda_1'/2)u^2(\eta)\Big].
            \end{split}
	    \end{equation}

	    Thus one can calculate the total energy for the PM state by simply integrating the energy density,
	
	    \begin{equation}\label{eq:rev70}
    	    E_{PM}(\eta_0)=\int_{0}^{\eta_0}d\eta \eta^2 e_{PM}(u(\eta))+\int_{\eta_0}^{\infty}d\eta \eta^2 e_{PM}(u(\eta,\eta_0)).
	    \end{equation}
	    Now the minimization of energy with respect to the matching point $\eta_0$ gives the energy of PM state.
        
    \subsection{Effect of the variational method}
        For a spin-1 BEC with anti-ferromagnetic type of interaction in a 3-D harmonic confinement, following T-F approximation the relative energy difference between the PM and the polar( or APM or AF) state ($(E^{TF}_{PM}-E^{TF}_{pol})/E^{TF}_{PM}$) is found out to be $\simeq0.3\%$ with polar state winning energetically at $p=0$ and $q=0$. This energy difference is invariant of the number of condensate particles. So our main purpose to employ variational method (VM) is to see its effect on the energy difference as well as its dependence on the number of the condensate particles. The energy difference being very small as predicted by the T-F result, there is a possibility that the inclusion of the correction might favor the PM state to win energetically. 
        \begin{table}[htbp]
		    {\begin{tabular}{ |m{1.8cm}|m{3cm}|m{3cm}| }
				\hline
				&\multicolumn{2}{|c|}{} \\
				Particle Number & Polar/APM/AF &PM state\\
				\hline
				$N=1000$& $E^{TF}=3.0267$,\quad $E^{var}=4.8742$&$E^{TF}=3.0364$, $E^{var}=4.8807$ \\  
				\hline
				$N=5000$& $E^{TF}=28.8084$\quad $E^{var}=34.994$&$E^{TF}=28.9011$, $E^{var}=35.0768$  \\
				\hline
			    $N=10000$& $E^{T-F}=76.0257$,\quad$E^{var}=86.2813$ &$E^{TF}=76.2705$,\quad$E^{var}=86.5118$  \\
				\hline
				$N=15000$& $E^{TF}=134.1184$,\quad $E^{var}=147.8625$&$E^{TF}=134.5502$, $E^{var}=148.2776$ \\
				\hline
			\end{tabular}
    		}
		    \caption{Some of the numerical values of the total energy for different number of particles are shown. The energy scale of $3\hbar \omega/(2\lambda_0)$ is used, multiplying this scale with the above given energy values would produce the actual energy for the stationary states. Here, $E^{var}$ denotes the total energy obtained from VM and $E^{TF}$ is the T-F approximated total energy. All the values given are mostly rounded off in the last decimal places.}
        \end{table}
        
        \par
        
        So we need to compare the total energy of these two states taking into account the full profile of the condensate. But before calculating the energy we need to compute the parameter $\mu'$ first. This $\mu'$ can be calculated numerically (up to a specified accuracy level) for different values of $\eta_0$ using Eq.(\ref{eq:rev59}) (for polar state) and Eq.(\ref{eq:rev68}) (for PM state) for a given number of condensate particles. Once we know the $\mu'$ for different values of $\eta_0$, the number densities for $\eta\leq\eta_0$ and $\eta\geq\eta_0$ can be easily found out. As a result the energy density can be integrated to compute the total energy for different values of $\eta_0$. The free parameter $\eta_0$ can be fixed from the energy minimization condition for both the states. Note that, $\mu'$ is computed for varying free parameter $\eta_0$ for a particular condensate particle $N$. Once $\eta_0$ is fixed from the energy minimization, $\mu'$ is also accurately found out.
        For wide range of condensate particles in an experimentally achievable regime (e.g. $N=1000,3000,5000,7000,10000,15000$) we employ the VM for the 3-D isotropic harmonic confinement, where we stick to the trapping frequency, $\omega=2\pi\times 100\quad Hz$. The interaction parameters are, $\lambda_0=3.88164\times 10^{-3}$ and $\lambda_1'=0.0161$. Some numerical values of energy corresponding to the PM and the polar state is summarized in Table-$1$.

        \begin{figure}[h!]
    		\subfloat[$\eta_0/\eta_{T-F}\quad vs \quad N$\label{subfig-1:matching point}]{%
    			\includegraphics[width=0.46\textwidth]{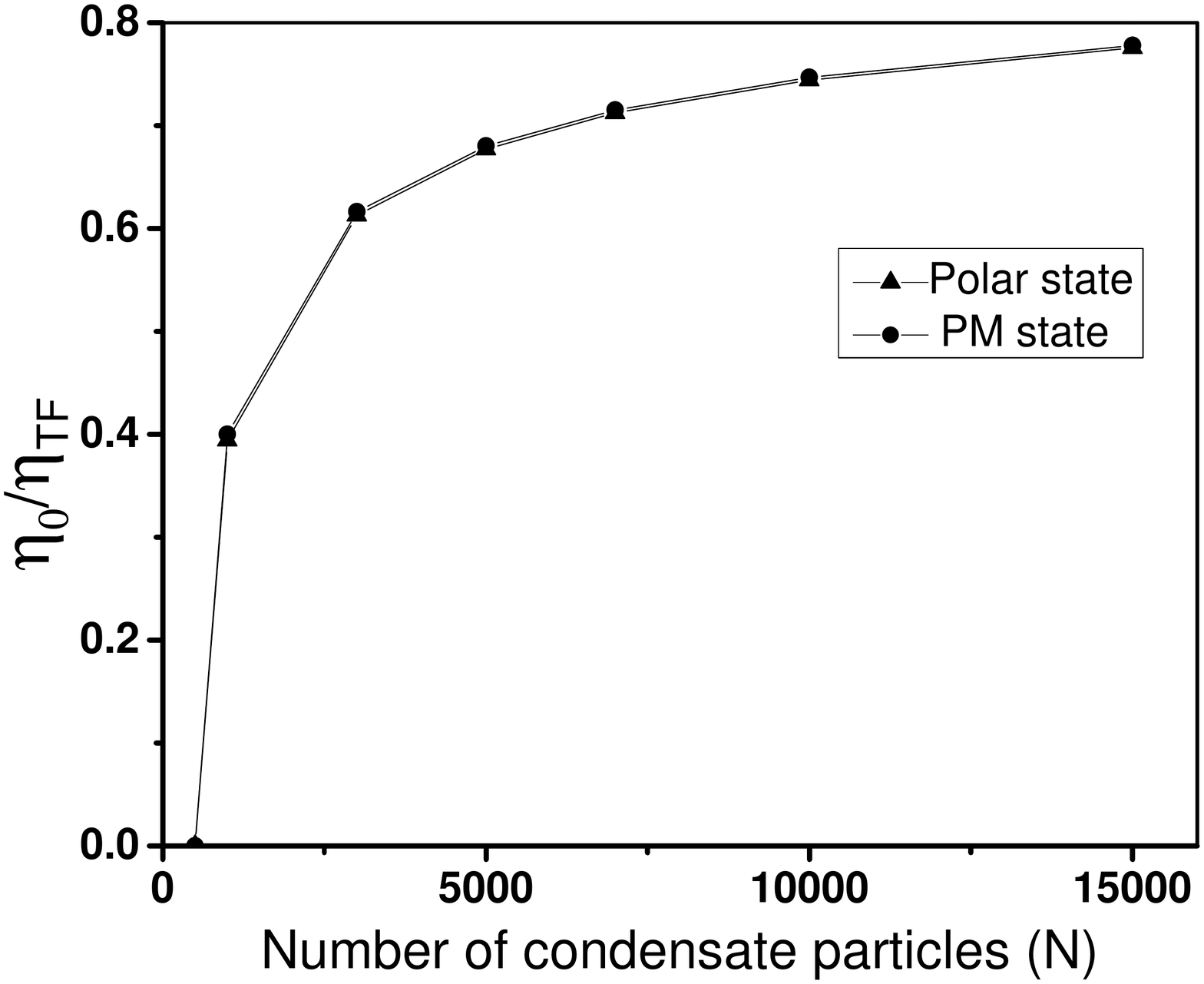}
    		}
    		\\
    		\subfloat[$\mu'/\mu'_{T-F}\quad vs\quad N$\label{subfig-2:chemical potential}]{%
    			\includegraphics[width=0.46\textwidth]{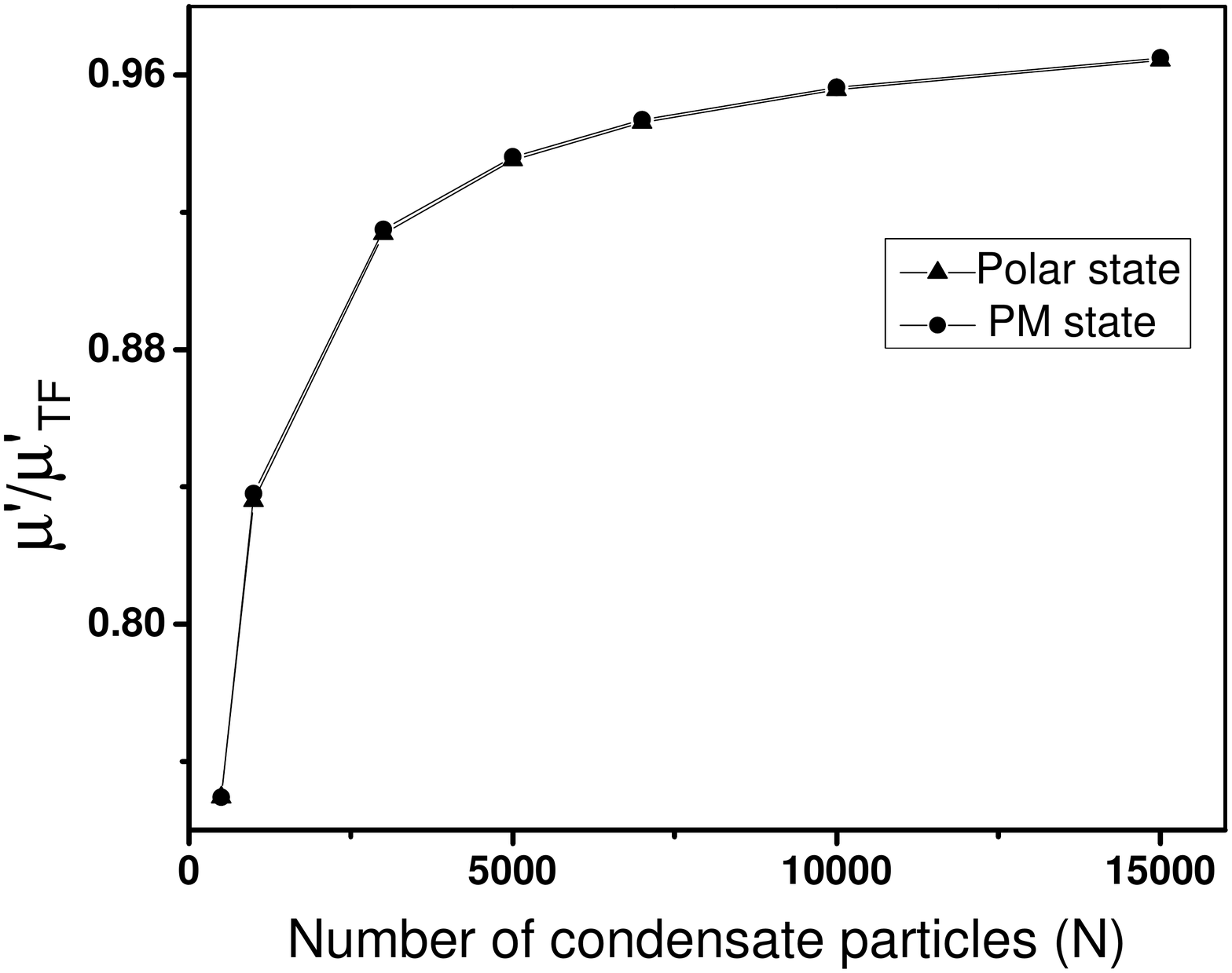}
    		}
    		\caption{Subfig.-(a): The ratio of $\eta_0$ and the T-F radius varies with number of condensate particles. This ratio also tells about the accuracy of the T-F results. As the number of condensate particles is increased the matching point shifts towards the T-F radius for both polar and PM state. Thus the effect of correction becomes more and more irrelevant. But the VM is more relevant when the condensate particles are less than 10,000, given our choice of the parameter values ($\omega$, $\lambda_0$ and $\lambda_1'$). Subfig.-(b): The ratio of $\mu'$ obtained via the VM and the T-F approximation increases with the number of condensate particles. the number density at the center of the trap is proportional to $\mu'$ with the same proportionality constant for both VM and T-F. So, this ratio also produces the number density at the core of the trap ($\eta$=0) estimated via VM and T-F. This ratio approaches to unity as the condensate becomes larger but for lower number of condensate particle the $\mu'$ obtained from VM deviates significantly from that of T-F approximation.}
    		\label{fig:eta and mu}
    	\end{figure}
        \begin{figure}[h!]
    		\subfloat[$E_{Var}/E_{TF}\quad vs\quad N \quad in \quad 3D$\label{subfig-1:Evar_vsETF_3D}]{%
    			\includegraphics[width=0.46\textwidth]{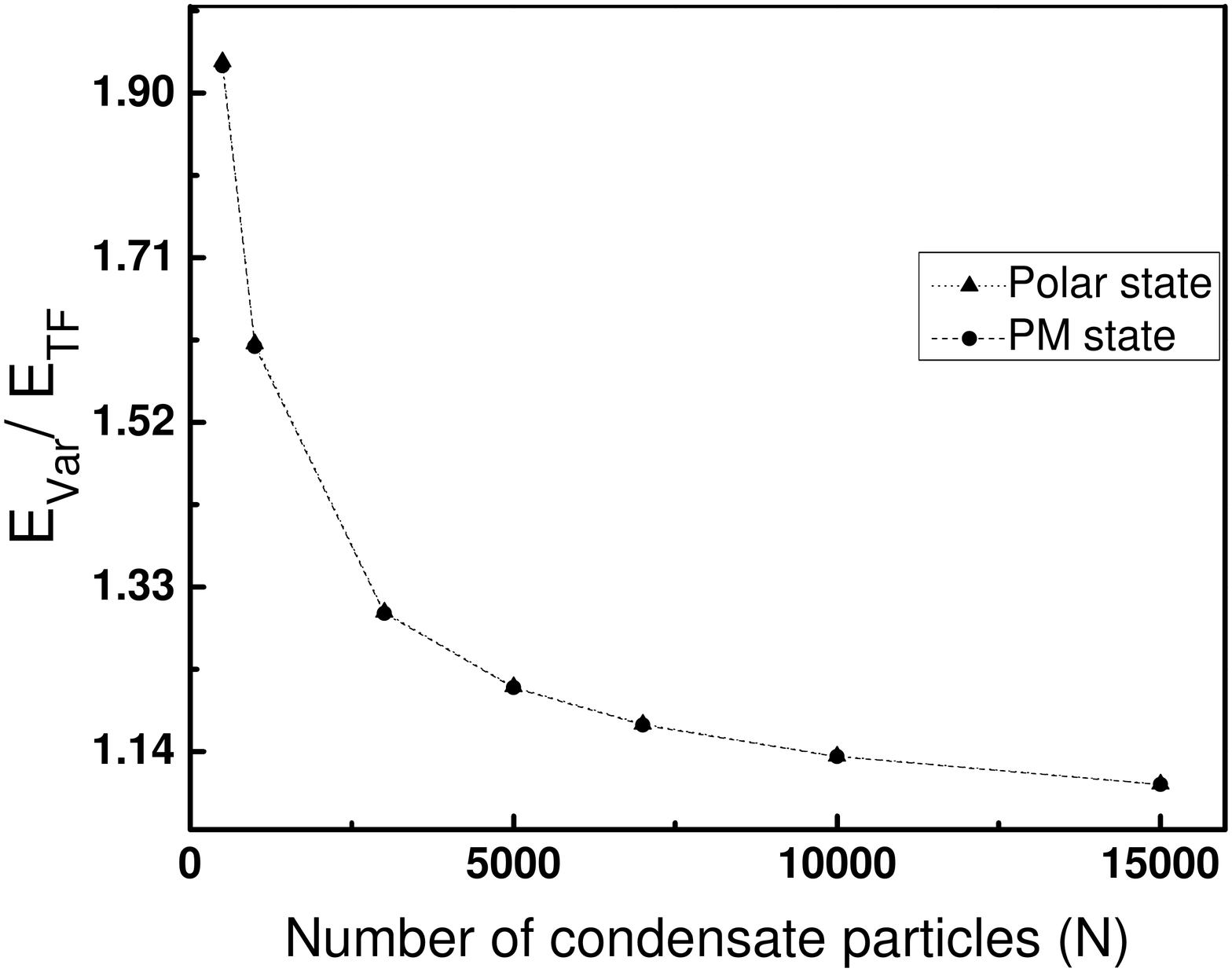}
        		}
    		\\
    		\subfloat[$\Delta E_{Var}/\Delta E_{TF}\quad vs\quad N \quad in \quad 3D$\label{subfig-2:energy_diff_3D}]{%
    			\includegraphics[width=0.46\textwidth]{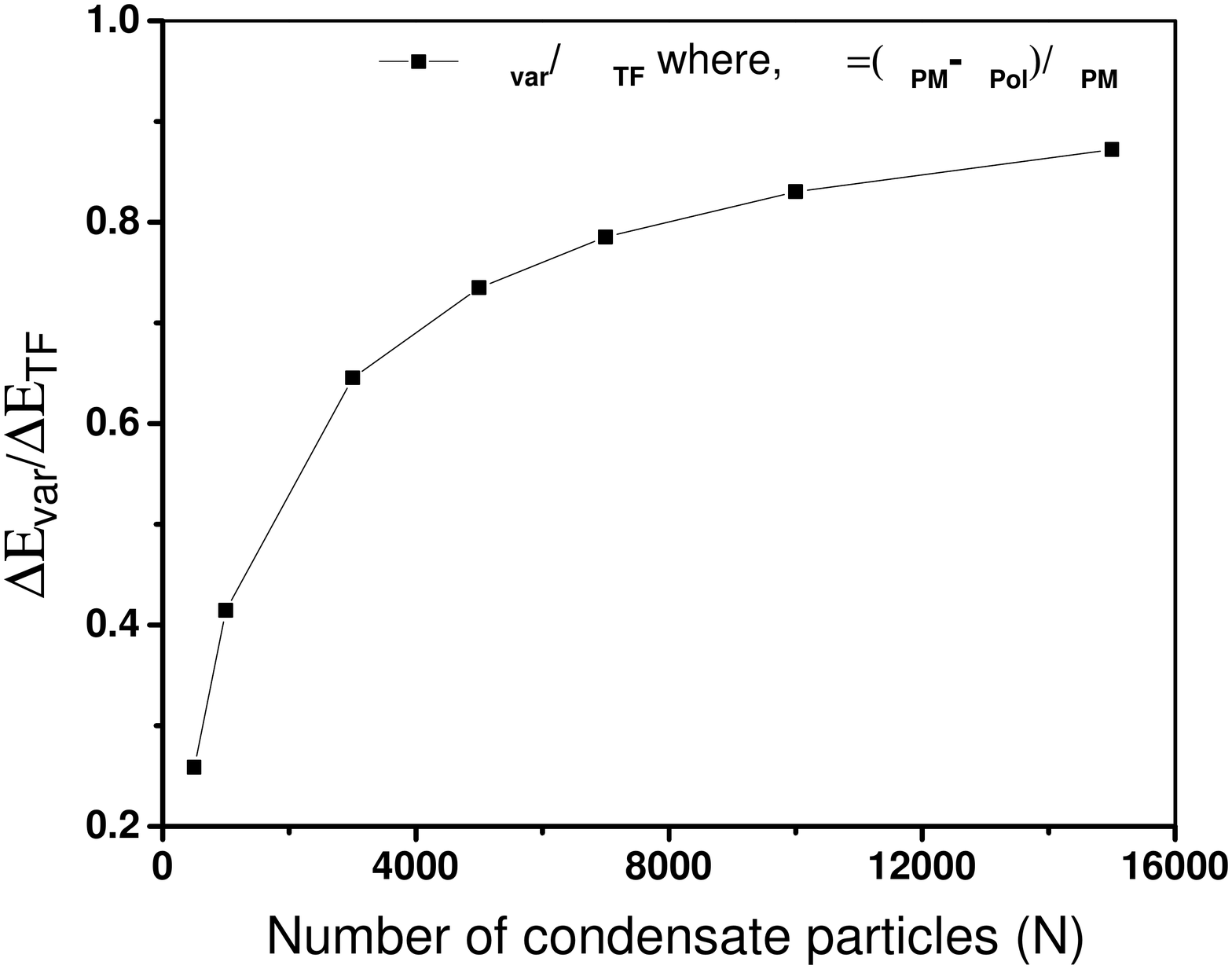}
        		}
    		\caption{Subfig.-(a) The ratio of total energy estimated by the VM and the T-F approximation for polar/APM/AF and PM states. The effect of VM on the energy is similar for both the stationary states, where for small condensates the energy ratio is almost twice. This depicts the significance of using VM instead of T-F approximation. Subfig.-(b): The ratio of variational and T-F approximated energy difference of PM state and polar state ($\Delta E=(E_{PM}-E_{Pol})/E_{PM}$) for different condensate particles clearly depicts that consideration of the full number density profile via variational approach does not change the fact that under harmonic trapping, polar/ AF/ APM state still energetically wins over the $PM$-state but inclusion of the correction actually reduces the energy difference between the polar and the PM state by a significant margin especially for the condensates with smaller number of particles.}
    		\label{fig:energy_diff_3D}
		\end{figure}
         It can also be concluded that the correction is relevant for the PM state as well as polar state as the energy estimated from the VM is significantly different from the T-F estimation (see appendix-$B$ for details of T-F in non-dimensional form). We also calculate the energies for N = 500, which is really small number of particles to observe the drastic difference between VM and T-F limit. 
         \par
         It is found out that $\eta_0$, the point where the Gaussian tail smoothly matches with the number density of inner region, moves toward the T-F radius as the number of condensate particles are increased meaning, for large condensates the T-F gives more accurate results (Fig.(\ref{fig:eta and mu}\subref{subfig-1:matching point})). A very similar trend is also followed by the $\mu'$ estimated in the VM (Fig.(\ref{fig:eta and mu}\subref{subfig-2:chemical potential})). For small condensate the VM estimated $\mu'$ is lesser than that of T-F estimation because of the significant contribution of the Gaussian tail. The ratio of the parameter $\mu'$ for VM and T-F also signifies the ratio of the number density (predicted by these two methods) at the centre of the trap, at $\eta=0$ (Fig.(\ref{fig:eta and mu}\subref{subfig-2:chemical potential})).  
         \par

        \begin{figure}[h!]
    		\subfloat[$E_{Var}/E_{TF}\quad vs\quad N \quad in \quad 1D$\label{subfig1-:Evar_ETF_1D}]{%
			    \includegraphics[width=0.46\textwidth]{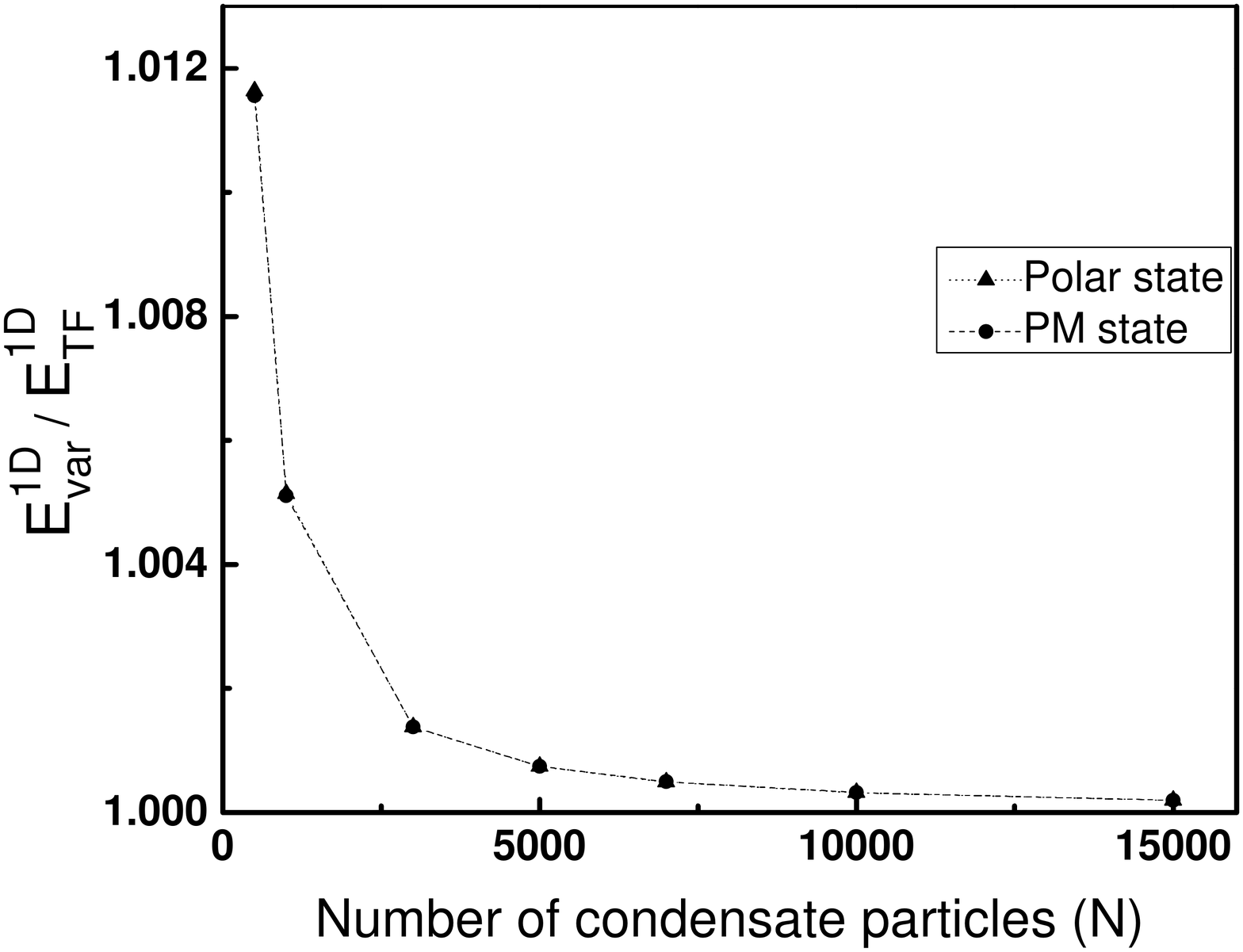}
        		}
    		\\
    		\subfloat[$\Delta E_{Var}/\Delta E_{TF}\quad vs\quad N \quad in \quad 1D$\label{subfig2-:energy_difference_1D}]{%
    			\includegraphics[width=0.46\textwidth]{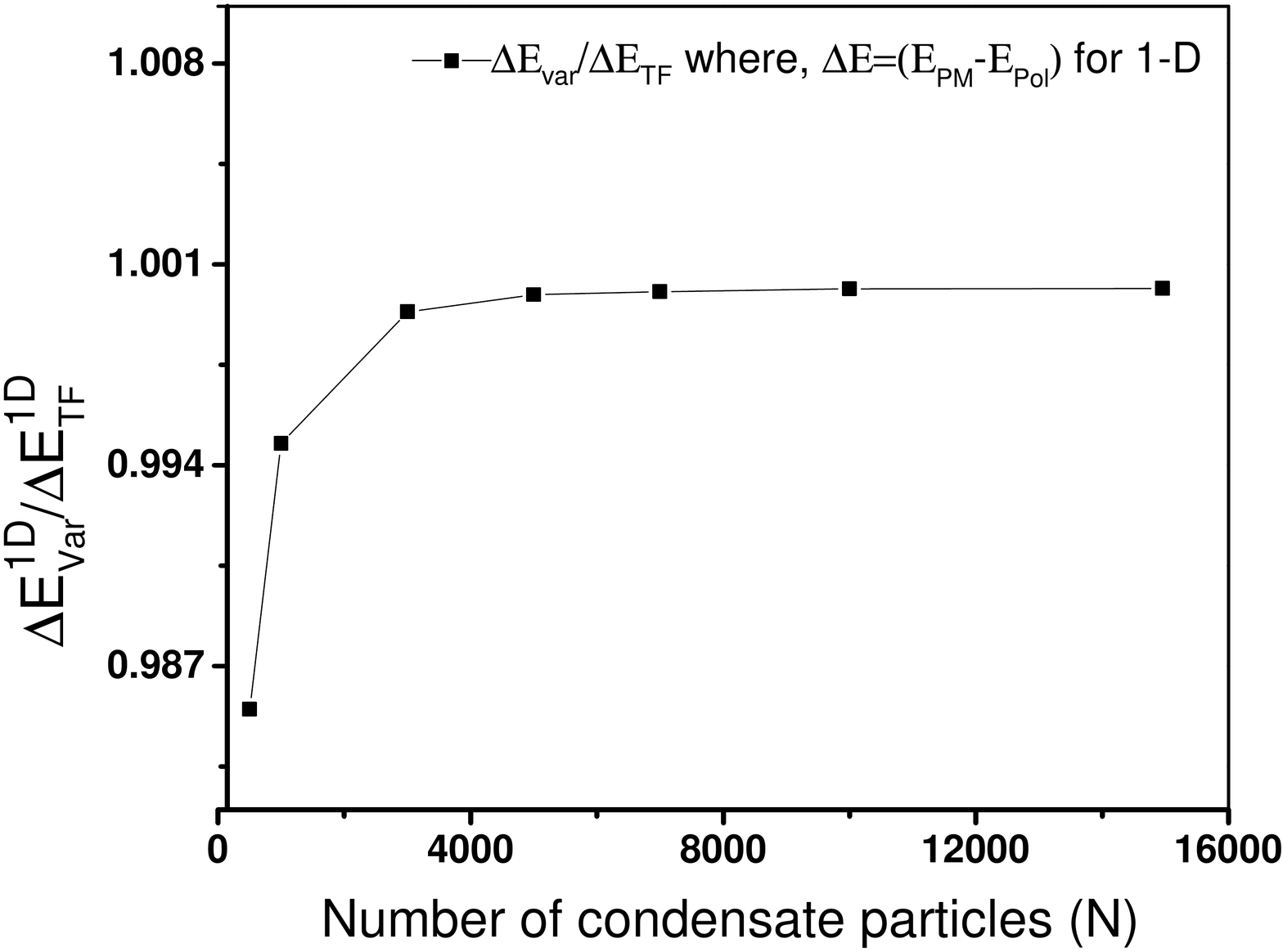}
        		}
    		\caption{Subfig.-(a): Ratio of total energy estimated by VM and T-F approximation varies with the total number of condensate particles. The energy estimated by the VM is always greater than the T-F limit, though for larger condensates the two method agrees really well. Even for small condensates of N = 500, 1000 the T-F approximation gives fairly accurate results for effective 1-D system (see Appendix-$A$ and Appendix-$B$ for details), given our choices of the parameter values. Subfig.-(b) The ratio of energy difference ($(E^{1D}_{PM}-E^{1D}_{pol})/E^{1D}_{PM}$) between the PM and polar/APM/AF states in VM and T-F is plotted with number of condensate particles which remains approximately unity even for smaller condensate under 1D harmonic confinement.}
    		\label{fig:energy_1D}
		\end{figure}
		
		Interestingly, we find that for this specific choice of trapping frequency (of the 3D harmonic trapping), the matching point is extremely closer to zero or might be equal to zero ($\eta_0\simeq 0$) for condensates with $N\leq 521$. Thus the VM predicts an analytical gaussian-like condensate profile. In next section, we will compare this predicted profile with computationally obtained condensate profile for $N=500$. 
		
		\par
        The VM is really significant especially for condensates with smaller particle number where we find that the total energy estimated for both PM and polar/APM/AF state is almost twice (N = 500) the energy estimated by the T-F limit (Fig.(\ref{fig:energy_diff_3D}\subref{subfig-1:Evar_vsETF_3D})).
        The energy difference between the polar and the PM state (which was $\simeq0.3\%$ in T-F limit) becomes almost $\simeq0.08\%$ for smaller condensates with N = 500 and $\simeq0.1\%$ for N = 1000, while for larger condensates too the energy difference is reduced in comparison to the T-F estimation though the polar state still becomes the ground state (Fig.(\ref{fig:energy_diff_3D}\subref{subfig-2:energy_diff_3D})).  
        \par
        The VM is also employed for 1-D condensate (for details see appendix-$A$), where we assume the geometric mean of the trapping frequencies in transverse direction $\omega_{yz}=2\pi\times 1261\quad Hz$ and The trapping frequency along the direction of elongation is $\omega_x=2\pi\times50 \quad Hz$. The corresponding interaction parameters are, $\lambda_0^{1D}=46.157075\times10^{-3}$ and $\lambda_1''=0.0161$ (all these parameters are defined in appendix-$A$). 
        The energy estimation in VM in effective 1-D trapping varies with the number of condensate particles in a similar fashion as the 3-D isotropic trapping where the VM is more pronounced for smaller condensates. But in comparison with the 3-D case (Fig.(\ref{fig:energy_diff_3D}\subref{subfig-1:Evar_vsETF_3D})) for this choice of trapping frequencies the energy estimated by VM does not deviate significantly (Fig.(\ref{fig:energy_1D}\subref{subfig1-:Evar_ETF_1D})) from the T-F approximated energy for both the stationary states. As a result the energy difference between the stationary states (which is $\simeq0.5\%$ in T-F limit for 1-D condensate) almost remains the same even for the smaller condensates even after considering the full profile of the condensate. For different choices of trapping frequencies more noticeable energy difference could be seen.
        
        \subsection{Comparison of variational method with numerically obtained condensate profile}  
        
        When the spin-spin interaction is of anti-ferromagnetic type, the variational method predicts that the degenerate states would be the ground state at zero magnetic field. But the variational method is done using a reasonable assumption of the number density at low density region. So, we start with the GP equation for polar state and numerically simulate the polar state condensate profile and compare it with the analytical profile to check the accuracy of the variational method.
        \par
        For condensates in 3-D isotropic harmonic trapping, we substitute
        $u\rightarrow |\phi(\eta)|^2/\eta^2$ , which allows for tackling the kinetic term efficiently \cite{Edwards_Burnett}. Here, we use the imaginary-time split-step Fourier method to find the ground state condensate profile. 
        
        \begin{figure}[h!]
    		\subfloat[$|\phi(\eta)|^2=\eta^2\times u(\eta)$ vs radial distance ($\eta$) in  3-D for N=500\label{subfig1-:phi2_500}]{%
			    \includegraphics[width=0.46\textwidth]{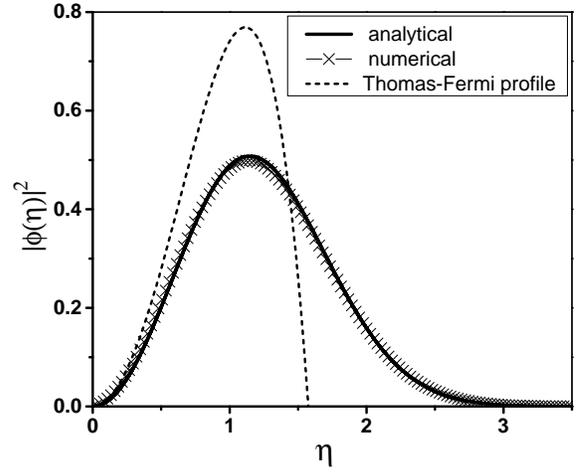}
        		}
    		\caption{The condensate profile represented in terms of $|\phi(\eta)|^2$. This $|\phi(\eta)|^2$ is related to the number density of the polar state as, $|\phi(\eta)|^2=\eta^2\times u(\eta)$. The analytical profile from VM matches quite well with the numerically simulated condensate profile even for small condensates with particle number as low as 500. It is well known that the Thomas-Fermi approximation is typically valid for large condensates. No surprise that the Thomas-Fermi profile shown in dashed line deviates significantly with the simulated profile.}
    		\label{fig:profile_3D}
		\end{figure}
		
		\begin{figure}[h!]
    		\subfloat[number density $u(\zeta)$ with distance $\zeta$ for N=500\label{subfig1-:1D 500}]{%
			    \includegraphics[width=0.46\textwidth]{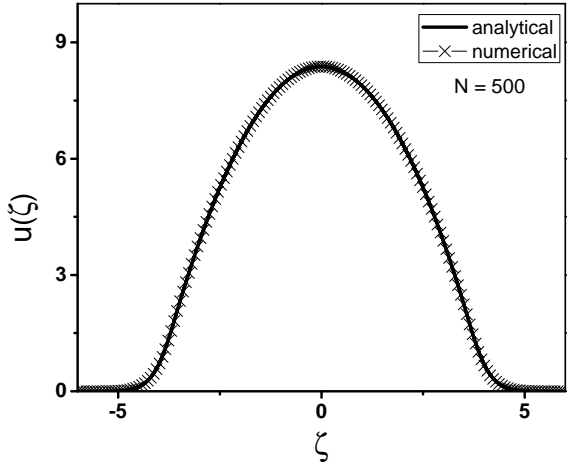}
        		}
    		\\
    		\subfloat[number density $u(\zeta)$ with distance $\zeta$ for N=15000\label{subfig2-:1D 15000}]{%
    			\includegraphics[width=0.46\textwidth]{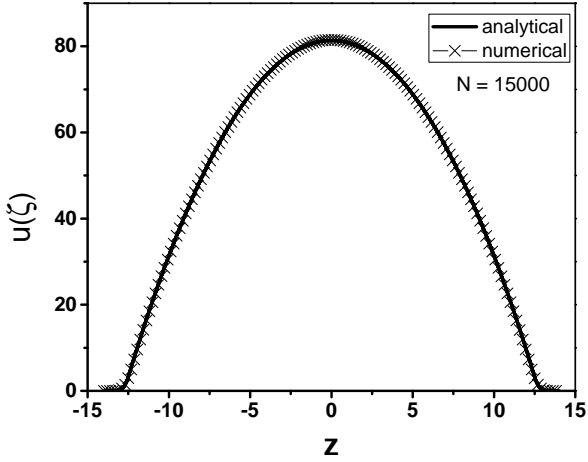}
        		}
    		\caption{Numerically obtained number density profile is compared with the analytical profile presented here. Subfig.-(a): For small condensate with particle number as low as N= 500, the analytical profile with $\mu'\simeq$ 8.381 and the matching point at ($\zeta_0\simeq$) $\pm$3.611 agrees quite well with the numerically obtained profile. Due to different length scale specific to 1-D confinement the spatial distance is denoted as $\zeta$ (see Appendix-A for details). Subfig.-(b) For large condensate (N = 15000) we find $\mu\simeq$  81.373 and $\zeta_0\simeq$ 12.45. Typically for large condensate the T-F approximation is a very good approximation as a result we find the VM approximated $\mu'$ is close to the T-F approximated $\mu'$ and the matching point is not very far from the T-F radius. We find the analytical profile matches with the numerical profile quite well.}
    		\label{fig:profile_1D}
		\end{figure}
	
		\par
        For small condensates with 500 particles we find $\eta_0\simeq 0$ from the variational method. So the analytical number density profile is $u=(\mu'+\eta^2/2)exp(-\eta^2/\mu'))$  corresponding to, $\mu'=0.93$. In Fig.(\ref{fig:profile_3D}\subref{subfig1-:phi2_500}) the 3-D condensate profile is represented in terms of $|\phi(\eta)|^2$ where, the analytical profile is $|\phi^{Anal.}(\eta)|^2=\eta^2(\mu'+\eta^2/2)exp(-\eta^2/\mu'))$.
        \par
        We see that the analytical profile predicted by the variational method aptly represents the numerically simulated profile with reasonable accuracy where the T-F profile $|\phi^{T-F}(\eta)|^2=\eta^2(\mu'^{T-F}-\eta^2/2)$
        with the T-F approximated chemical potential, $\mu'^{T-F}=1.241$, fails as expected.
        \par
		For the 1-D harmonic trapping the variationally obtained number density of the polar state is compared with the numerically obtained density profile for both small (Fig.(\ref{fig:profile_1D}\subref{subfig1-:1D 500})) and large (Fig.(\ref{fig:profile_1D}\subref{subfig2-:1D 15000})) condensates. We find that the numerically obtained profile is in excellent agreement with the analytical profile in both the cases.

		We observe that the matching point ($\eta_0$) depends on $\lambda_0N$. The parameter $\lambda_0$ can be modulated via Feshbach resonance (via changing $c_0$) or by changing the trapping frequency (hence the oscillator length). Thus for a 3-D confinement, if the parameter $\lambda_0$ is halved, the matching point ($\eta_0$) is found to be close to zero or almost zero ($\eta_0\simeq0$) even for larger condensate with $N$ = 1000. Thus given such scenario even for larger condensates, the VM method would find its significance for providing a fairly accurate condensate profile.
		
		\subsection{Energy Density of Polar and PM state:}
\begin{figure}[h!]
    		\subfloat[energy density $e(\eta)$ with distance $\eta$ from the centre of the trap for N=1000\label{subfig1-:1D energy density 1k}]{%
		    \includegraphics[width=0.46\textwidth]{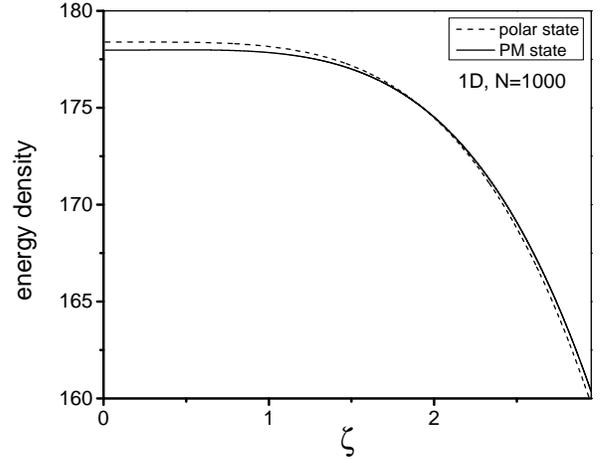}
    		}
    		\caption{The energy density for polar and PM state is plotted against the distance ($\zeta$), where $\zeta=0$ is the 1-D trap center. The PM state has lower energy density near the center of the trap while if we move away from the trap center the polar state has lower energy density. So there is a possibility that PM state in the core of the trap and polar state in the outer region would further decrease the total energy. Thus, there is a possibility of domain formation in absence of the Zeeman terms i.e. $p=0$ and $q=0$. The possible domain boundary can also be predicted from this plot, approximately at $\zeta\simeq1.93$.}
    		\label{fig:energy_density_1D}
		\end{figure}

		As the variational method we are employing is of multi-modal in nature and analytically estimates the number density with excellent accuracy, this method can also be employed to analytically explore the phase-separation phenomenon.
		Here the energy density for the energetically close stationary states are plotted against the radial distance from the trap center for the 1-D harmonic trapping Fig.~\ref{fig:energy_density_1D}.
		\par
		Though the total energy for the three degenerate states (polar, APM and AF) is lesser than the PM state, we observe that in the core region (in the trap center) the energy density of the PM state is lesser than the degenerate states. So according to VM, the total energy can be further lowered if the PM state is at the central region and the degenerate states are at the peripheral region.
		
		\par
		We compute the spin-1 system for zero magnetic field and zero magnetization. We start with the PM state at the center and the degenerate states outside as initial conditions and use imaginary time propagation to get to the ground state \cite{bao_numerical,gautam_GPE_solver}. After long time propagation we observe that the degenerate states become the ground state and no phase separation arises. This might be due to the domain wall energy which probably increases the total energy.
		\par
		But importantly, we find that depending on the initial condition any one of the degenerate states becomes the ground state which is natural, whereas, in \cite{bao_numerical,gautam_GPE_solver} only AF state becomes the ground state at zero magnetic field and zero magnetization for and anti-ferromagnetic condensate.
		\par
		It is straightforward to extend this variational method for spin-1 system in presence of magnetic field. So the analytical method presented here would be of general importance to understand different phase separation possibilities and when the domain wall energy comes into play even in presence of magnetic field.
		\begin{widetext}
		\begin{figure*}[ht]
    		\subfloat[initial condition: PM state in the core with AF outside.\label{subfig1-:1D AF}]{%
			    \includegraphics[width=0.33\textwidth]{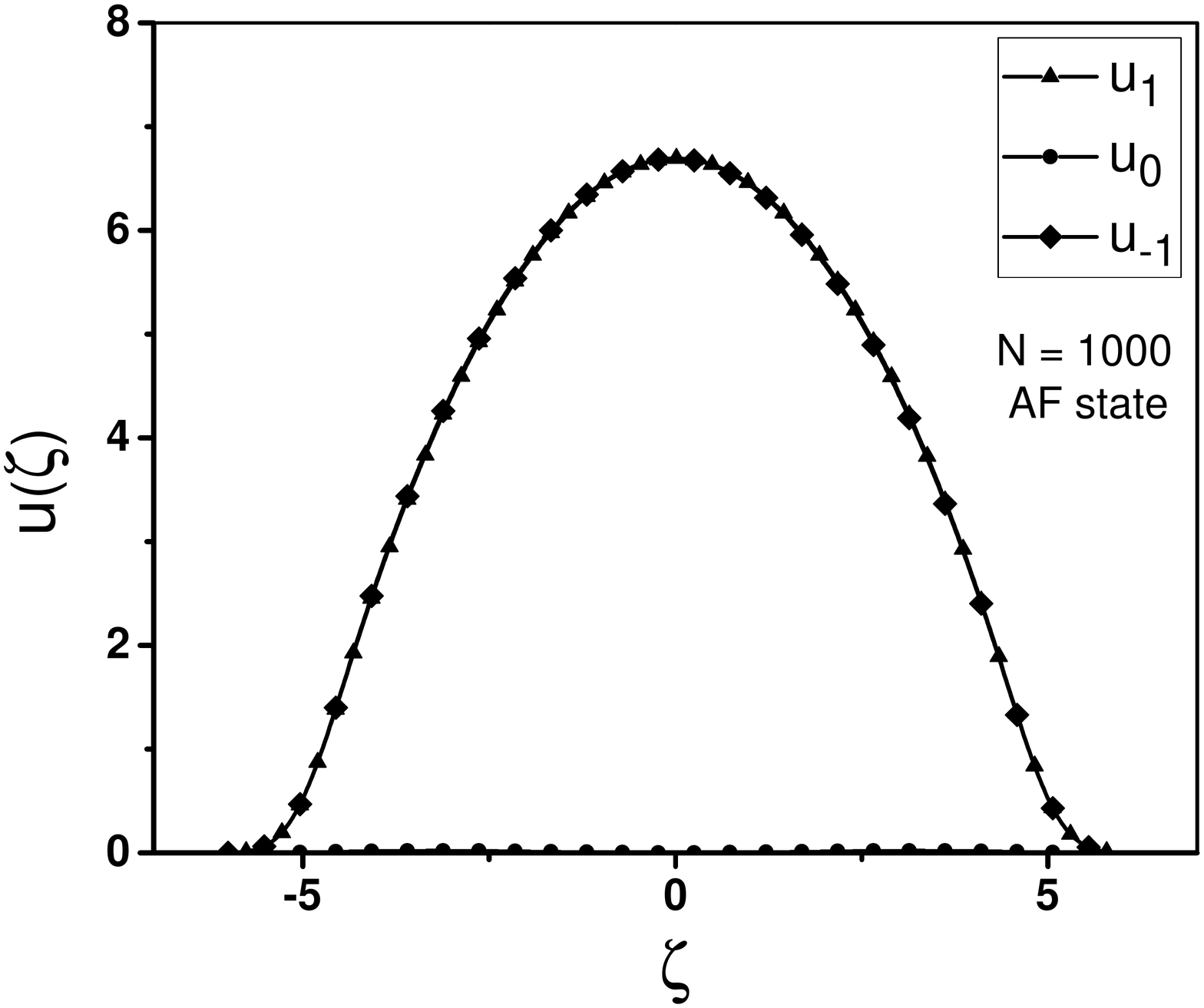}
        		}
    		\subfloat[initial condition: PM state in the core with APM outside.\label{subfig2-:1D APM}]{%
    			\includegraphics[width=0.33\textwidth]{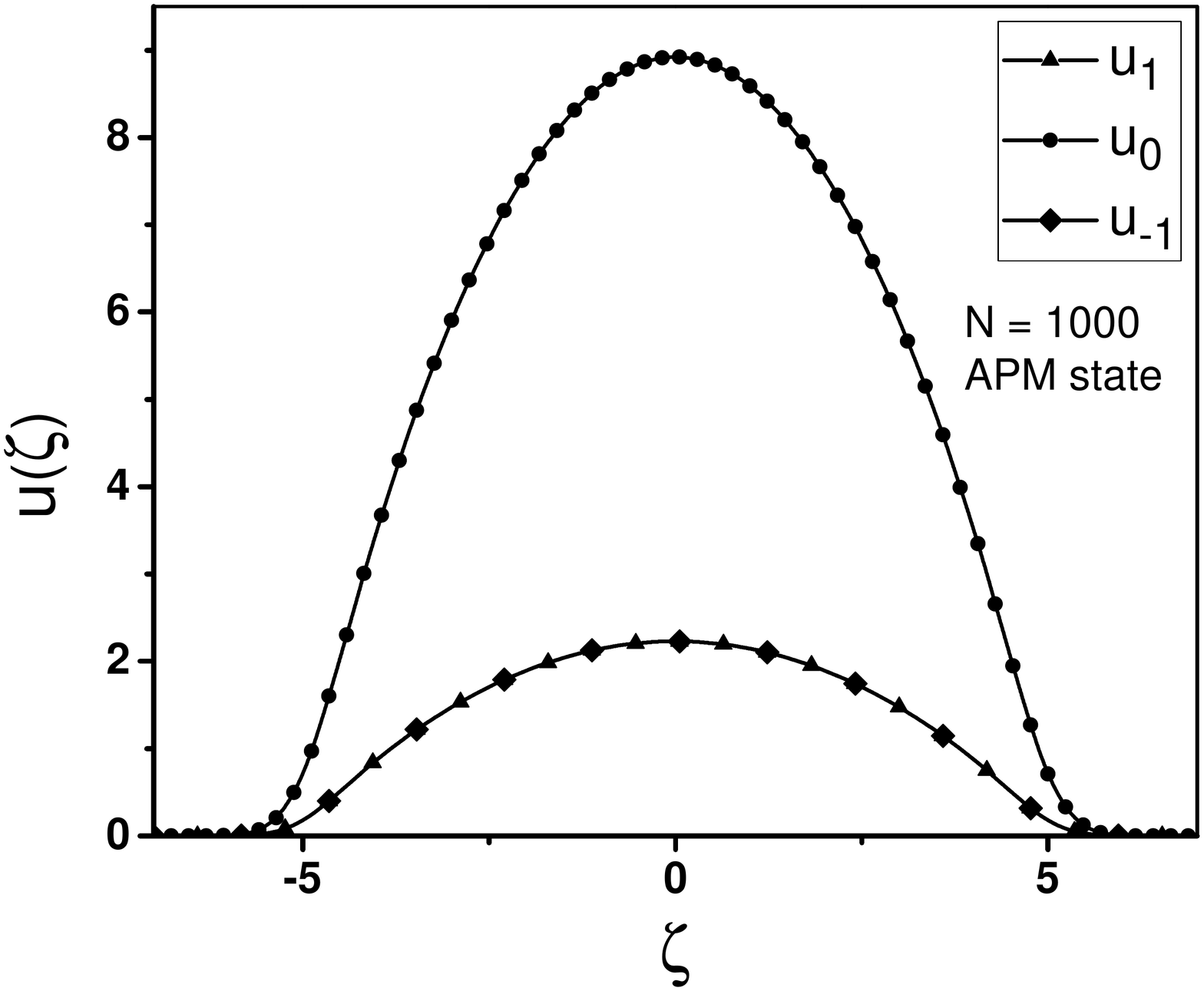}
        		}
    		\subfloat[initial condition: PM state in the core with polar outside.\label{subfig2-:1D polar}]{%
    			\includegraphics[width=0.33\textwidth]{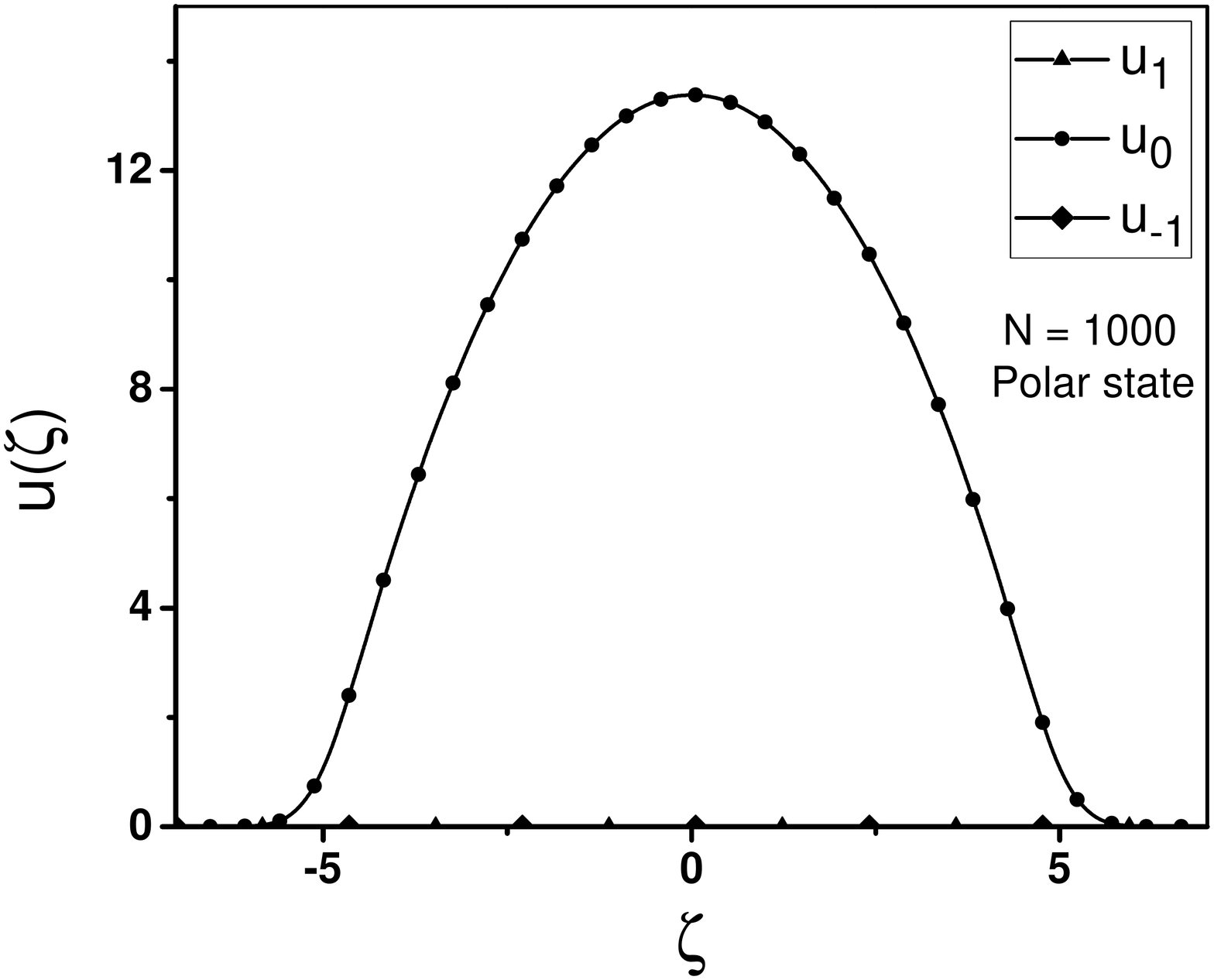}
        		}
    		\caption{We start with the VM prediction of PM state near the centre of the trap for the effective 1D trap followed by the degenerate states i.e. (a) anti-ferromagnetic, (b) anti-phase-matched and (c) polar state at a distance roughly $\zeta\simeq$1.93 (Fig.\ref{fig:energy_density_1D}) as initial conditions. After imaginary time propagation for adequate amount of steps the solution converges to the degenerate states taken in the peripheral region with no PM state.}
    		\label{fig:spin_profile_1D}
		\end{figure*}
		\end{widetext}
    
    
    \section{Discussion}
    In  this article we have introduced a variational method which is multi-modal in nature and estimates the number density profile of the spin-1 BEC fairly accurately in comparison with numerical profile. To use this method we choose the simplest context of ground state structure of a spin-1 BEC in absence of magnetic field. 
    \par
     First, we have looked into the ground structure in details for a spin-1 condensate by employing the T-F approximation for both ferromagnetic and anti-ferromagnetic type of spin-spin interaction. In T-F limit the ferromagnetic state becomes the ground state for ferromagnetic type of interaction. For anti-ferromagnetic type of interaction, polar, anti-ferromagnetic and anti-phase-matched state are equally likely to be the ground state. In T-F limit we see that the energy difference of these degenerate states with the closely competing phase-matched state is very small therefore, there is a chance that considering the full number density profile instead of the T-F profile may change the energy difference and may favor the phase-matched state as the ground state, especially for condensates with fewer particles where T-F approximation fails. So to investigate the situation further, we use our variational method which considers the full number density profile of the condensate and works for every mode. This technique also preserves the degeneracy of the polar, anti-ferromagnetic and the anti-phase-matched state at zero field.
    \par
    The new method of variation that we propose in this paper is constructed to produce a smooth number- and energy-density profile for even a few hundred of condensate particles in harmonic confinement. Though it still predicts that the degenerate states will be the ground state at zero field when the spin interaction is anti-ferromagnetic type, it changes the total energies of the stationary states substantially, especially for condensates with smaller number of particles in a 3-dimensional isotropic harmonic trapping. The variational method also predicts that this relative energy difference depends on the total number of condensate particles (N), whereas in T-F approximation, it is independent of $N$. For small condensates, this relative energy difference is decreased by a factor of $\simeq$ 1/4. We also investigate the scenario of spin-1 BEC confined in an effective 1-dimensional trapping in zero field. The variational method does not change the total energy and the relative energy difference between the PM and the degenerate states by any substantial margin for the specific choice of parameter values. Hence, for the specific choice of the trapping potentials, it can also be concluded that the T-F limit works well for effective 1-D condensates even for small particle number whereas, for 3-D confinement variational method is highly significant (in comparison with T-F) even for large condensates. 
    \par
    We find that the analytical profile of the condensate as predicted by VM is fairly accurate in comparison to numerically obtained profile. Especially for condensates with small number of particles, the T-F profile is not at all accurate, whereas the VM profile is in excellent agreement with the numerical number density profile (Fig.(\ref{fig:profile_3D}\subref{subfig1-:phi2_500})). 
    \par
    Since the VM also provides a smooth energy density profile where the kinetic energy contribution is also included, from this energy density comparison of different stationary states one can infer possibilities of phase separation. In this simplest setting of $p=0$ and $q=0$ we are considering here, we found that there is a possibility of having phase separation (Fig.(\ref{fig:energy_density_1D})). By solving the GP equation numerically, with PM state close to the trap center and one of the degenerate state outside as initial conditions we see no phase separation. Instead, we see that the degenerate state we had chosen in the initial condition becomes the ground state of the system. So polar or AF or APM are equally likely to be the ground state in this limit of anti-ferromagnetic interaction. This is in contrast to the widely reported numerical result of AF state becoming the ground state of the system. We think the phase-separation is ruled out due to some other reason for example, the cost of domain wall energy. Nevertheless, this VM presented here might be really useful as the first step for identifying whether there is a phase-separation possibility as this method can be easily extended for spin-1 BEC in presence of magnetic field. The analytical profile produced by VM is also relevant even non-spinor BEC.
    
    \section{Acknowledgements}
    PKK would like to thank the Council of Scientific and Industrial Research (CSIR), India for providing funding during this research.

\bibliographystyle{apsrev4-1}
\bibliography{main} 	 
\appendix

\appendix

\section{Variational approach for 1-D condensate}
    
    For a quasi-one-dimensional condensate the interaction parameter and the number densities can be scaled as,
    \begin{equation}\label{eq:A1}
  	    c_0=2 \pi l^2_{yz} l_x\lambda^{1D}_0\hbar\omega_x, \quad c_1=2 \pi l^2_{yz} l_x\lambda^{1D}_1\hbar\omega_x,
    \end{equation}
    \begin{equation}\label{eq:A2}
        u_m=2 \pi l^2_{yz} l_x \lambda^{1D}_0 n_m, \quad r=l_x\zeta
    \end{equation}
    where, $l_x^2=\hbar/(m\omega_x)$, $l_{yz}^2=\hbar/(m\omega_{yz})$, $\omega_{yz}=\sqrt{\omega_y\omega_z}$ and $N$ is the total number of particles in the condensate. As a result, the parameters $\lambda^{1D}_0$, $\lambda^{1D}_1$, $\zeta$ and $u_m$ become all dimensionless. Note that we are considering the condensate to be elongated in $x$ direction as the harmonic trapping is far lesser than the geometric mean of the trapping frequencies along the other two direction i.e. $\omega_x<<\sqrt{\omega_{xy}}$
    \par
	Thus, the phase-stationary GP equation can be rewritten in dimensionless form as,
    \begin{equation}\label{eq:A3}
	    \begin{split}
	        \bigg\{ -\dfrac{1}{2}\dfrac{d^2}{d\zeta^2}+&\dfrac{1}{2}\zeta^2+ u-\mu'\\
	        &+\lambda'_1\left(u_1+u_{-1}+2 \sqrt{u_{-1}u_1}\cos\theta_r\right) \bigg\} \sqrt{u_0}=0,
	    \end{split}
	\end{equation} 
	\begin{equation}\label{eq:A4}
		\begin{split}
		\bigg\{ -\dfrac{1}{2}\dfrac{d^2}{d\zeta^2}+&\dfrac{1}{2}\zeta^2+ u-\mu' \pm \lambda'_1\left(u_1-u_{-1}\right)\bigg\}\sqrt{u_{\pm1}}\\
		&+\lambda'_1u_0\left(\sqrt{u_{\pm1}} +\sqrt{u_{\mp1}}\cos\theta_r\right)=0.
		\end{split}
	\end{equation}
    
	where, $\lambda_1''=\lambda^{1D}_1/\lambda^{1D}_0$ and $\mu'=\mu/(\hbar \omega_x)$. The total number density $u$ is the sum of all the sub-component density, i.e. $u=u_1+u_0+u_{-1}$. 
	\par
	Note that the equations have similar form with those for 3-D except for the kinetic term and interaction parameters due to different scaling.
		
	\subsection{Method of variational approach for 1D:}
	Following the similar approach as of 3D condensate we assume the number densities,
 	\begin{enumerate}[\indent {}]
 		\item{$u^{in}(\zeta)=f(\mu',\zeta)$ for $|\zeta|<|\zeta_0|$},
 		\item{$u^{out}(\zeta)=(a+c|\zeta|+d\zeta^2) \exp(-\dfrac{\zeta^2}{b})$ for $|\zeta|\geq|\zeta_0|$};
 	\end{enumerate}
 	given, the number density and its derivatives on both sides should be equal at $\zeta=\zeta_0$. As discussed, the exact functional form $f(\mu',\zeta)$ is different for different stationary states and can be found from the solution of the Eq.(\ref{eq:A3}-\ref{eq:A4}) by neglecting the kinetic energy term. Considering the symmetry of the problem, only the positive values for $\zeta$ and $\zeta_0$ are taken in the subsequent analysis to simplify notations.

 	\subsubsection{Polar state:}
 	    
 	    Following the same procedure as in 3D condensate for polar state, the number density for the high density region is found out,
 	     \begin{enumerate}[\indent {}]
 	        \item{$u^{in}_{pol}(\zeta)=\mu'-\zeta^2/2$ for $\zeta<\zeta_0$}.
 	    \end{enumerate}
 	    From the smooth matching condition at $\zeta=\zeta_0$ we find all the coefficients $a$, $b$, $c$, $d$,
 	    \begin{enumerate}[\indent {}]
 	        \item $d =\exp{\Big(\dfrac{12 \zeta_0^2}{k}}\Big) \dfrac{(k-12\mu')\mu'-(k-20\mu')\zeta_0^2-13 \zeta_0^4 }{2 \zeta_0^2 (-2 \mu' + \zeta_0^2)},$
            \item $b = \dfrac{k}{12}$ 
            \item $c = 48 \zeta_0^3\exp{\Big(\dfrac{12 \zeta_0^2}{k}}\Big)  \dfrac{k-12(\mu'-\zeta_0^2/2)}{k^2},$
            \item $a = \exp{\Big(\dfrac{12 \zeta_0^2}{k}}\Big)\dfrac{
            42\zeta_0^4-4(k-14 \mu'+22\zeta_0^2)\mu'+3k\zeta_0^2}{4 (2 \mu' - \zeta_0^2)},
            $
            
 	    \end{enumerate}
 	    where, $k = 6 \mu' - 9 \zeta_0^2 + \sqrt{36 \mu'^2 - 12 \mu' \zeta_0^2 + 33 \zeta_0^4}$.
 	    To determine $\mu'$, we use the total number conservation condition,
 	    \begin{equation}\label{eq:A5}
 	        \int^{\zeta_0}_0 u^{in}_{pol} d\zeta+\int^{\infty}_{\zeta_0} u^{out}_{pol} d\zeta =\lambda^{1D}_0 N.
 	    \end{equation}
        Following the integration and simplifying further we get the equation,
        \begin{equation}\label{eq:A6}
            \begin{split}
                 &\dfrac{1}{16 \sqrt{3}k^2}\Bigg[4 \sqrt{3}\zeta_0 \bigg(12 k \mu'^2 + 4 \mu' (-5 k + 24 \mu') \zeta_0^2- 216 \zeta_0^6\\
                 &+ (-53 k + 384 \mu') \zeta_0^4 \bigg) - \exp{\Bigg(\dfrac{12 \eta_0^2}{k1}}\Bigg) \sqrt{k\pi}\bigg(-60 k \mu'^2 \\
                 &+ 20 (7 k - 24 \mu') \mu' \zeta_0^2 + (29 k - 576 \mu') \zeta_0^4 \\
                 &+ 408 \zeta_0^6\bigg) Erfc\Big(\dfrac{2 \sqrt{3} \zeta_0}{\sqrt{k}}\Big)\Bigg] + \mu' \zeta_0 - \zeta_0^3/6=\lambda^{1D}_0N,
            \end{split}
        \end{equation}
    where, $Erfc((2 \sqrt{3} \zeta_0)/\sqrt{k})$ is the complementary error function and $k$ is defined earlier. Here $\lambda^{1D}_0$ is different from the 3D condensate due to scaling factor. Thus, $\mu'$ can be estimated numerically for different values of $\zeta_0$ and $N$. Though the procedure is same but the equation for determining $\mu'$ is different from the 3D condensate.
    \par
     The polar-state energy density can be written in these dimensionless parameters as (using Eqs.(\ref{eq:A1}-\ref{eq:A2}) in  Eq.(\ref{eq:rev9})),
    \begin{equation}\label{eq:A7}
        e_{pol}(u(\zeta))=\dfrac{\hbar\omega_x }{2\lambda^{1D}_0}\Big(-\sqrt{u(\zeta)}\dfrac{d^2}{d\zeta^2}\sqrt{u(\zeta)}+\zeta^2u(\zeta)+u^2(\zeta)\Big).
    \end{equation} 
    The total energy for polar state can be found out by integrating the energy density,
    \begin{equation}\label{eq:A8}
		E_{pol}(\zeta_0)=\int_{0}^{\zeta_0}d\zeta e_{pol}(u^{in}(\zeta))+\int_{\zeta_0}^{\infty}d\zeta e_{pol}(u^{out}(\zeta)).
    \end{equation}
   
    \par
   Now from the minima of the total energy with respect to $\zeta_0$ fixes the total energy corresponding to the stationary state as well as the corresponding $\mu'$.
	\subsubsection{Phase-matched state}
	     Following a similar approach, the total density in the high density region ($\zeta<\zeta_0$) is written as,
	    \begin{equation}\label{eq:A9}
	        u^{in}= \dfrac{\mu'-\zeta^2/2}{(1+\lambda_1'')},    
	    \end{equation}
	    where the sub-component densities are 
	    \begin{equation}\label{eq:A10}
	        u^{in}_{\pm 1}=u^{in}/4 \qquad u^{in}_{0}=u^{in}/2.
	    \end{equation}
	    In the low density region where the kinetic term plays significant role, the sub-component densities are,
	    \begin{equation}\label{eq:A11}
	        u_{\pm1}^{out}(\zeta)=\dfrac{(a+c\zeta+d\zeta^2)}{4(1+\lambda_1'')} \exp\left(-\dfrac{\zeta^2}{b}\right), 
	    \end{equation}
	    \begin{equation}\label{eq:A12}
	        u_{0}^{out}(\zeta)=\dfrac{(a+c\zeta+d\zeta^2)}{2(1+\lambda_1'')} \exp\left(-\dfrac{\zeta^2}{b}\right), 
	    \end{equation}
	    for $\zeta\geq\zeta_0$, which follows from the same smooth matching condition. The parameters $a$, $b$, $c$ and $d$ has the same expressions as shown in case of polar state, but the $\mu'$ and the matching point $\zeta_0$ would be different for the PM state.
	    Total number conservation for this stationary state in 1D leads to,
	    \begin{equation}\label{eq:A13}
            \begin{split}
                &\dfrac{1}{16 \sqrt{3}k^2}\Bigg[4 \sqrt{3}\zeta_0 \bigg(12 k \mu'^2 + 4 \mu' (-5 k + 24 \mu') \zeta_0^2- 216 \zeta_0^6\\
                 &+ (-53 k + 384 \mu') \zeta_0^4 \bigg) - \exp{\Bigg(\dfrac{12 \eta_0^2}{k1}}\Bigg) \sqrt{k\pi}\bigg(-60 k \mu'^2 \\
                 &+ 20 (7 k - 24 \mu') \mu' \zeta_0^2 + (29 k - 576 \mu') \zeta_0^4 \\
                 &+ 408 \zeta_0^6\bigg) Erfc\Big(\dfrac{2 \sqrt{3} \zeta_0}{\sqrt{k}}\Big)\Bigg] + \mu' \zeta_0 - \zeta_0^3/6\\
                 &\hspace{13em}=(1+\lambda_1'')\lambda_0^{1D}N.
            \end{split}
        \end{equation}
         The parameter $\mu'$ of the PM state can be computed from this equation for different values of $N$ and $\zeta_0$.
	
	    The energy density for PM state for 1-D harmonic confinement is,
        \begin{equation}\label{eq:A14}
            \begin{split}
            e_{PM}=\dfrac{\hbar\omega_x}{2\lambda^{1D}_0}&\Big[-\sqrt{u(\zeta)}\dfrac{d^2}{d\zeta^2}\sqrt{u(\zeta)}\\
            &+\zeta^2u(\zeta)+(1+\lambda_1''/2)u^2(\zeta)\Big]. \end{split}
	    \end{equation}
        
	    Now the minimization of the total energy (integrating the energy density) with respect to $\zeta_0$ helps to fix the parameter $\zeta_0$ and the total energy of the PM state. 
        
    \section{Energy calculation using T-F approximation in dimensionless form}
            
        For an isotropic harmonic 3-D confinement, its easy to see from Eq.(\ref{eq:rev52}-\ref{eq:rev53}) that the polar state the number density varies as,
        \begin{equation}\label{eq:B1}
            u^{TF}_{pol}=\mu^{TF}_{pol}-\eta^2/2.
        \end{equation}
        Similarly, for PM state, the total number density can be written as,
        \begin{equation}\label{eq:B2}
            u^{TF}_{PM}=\dfrac{\mu^{TF}_{PM}-\eta^2/2}{(1+\lambda_1')}.
        \end{equation}
        Where, the T-F chemical potential can be written in terms of the corresponding T-F radius,
        \begin{equation}\label{eq:B3}
            \mu^{TF}_{PM(pol)}=\dfrac{1}{2}(\eta^{TF}_{PM(pol)})^2
        \end{equation}
        Now from the conservation of total number of condensate particles,
        \begin{equation}\label{eq:B4}
            \int^{\eta_{PM(pol)}^{TF}}_0 u^{TF}_{PM(pol)}\eta^2 d\eta =\lambda_0 N/3,
        \end{equation}
        it is easy to get to the T-F radius. Thus the T-F energy can be calculated easily.
            
        Similarly for 1D condensate,
        \begin{equation}\label{eq:B5}
            u^{TF}_{pol}=\mu^{TF}_{pol}-\zeta^2/2.
        \end{equation}
        \begin{equation}\label{eq:B6}
            u^{TF}_{PM}=\dfrac{\mu^{TF}_{PM}-\zeta^2/2}{(1+\lambda_1'')}.
        \end{equation}
        \begin{equation}\label{eq:B7}
            \mu^{TF}_{PM(pol)}=\dfrac{1}{2}(\zeta^{TF}_{PM(pol)})^2
        \end{equation}
        To get the 1-D T-F radius we can use the total number conservation equation which for 1-D can be written as,
        \begin{equation}\label{eq:B8}
            \int^{\zeta_{PM(pol)}^{TF}}_0 u^{TF}_{PM(pol)} d\zeta =\lambda_0 N.
        \end{equation}
        The total energy in dimensionless form 1-D in polar state,
        \begin{equation}\label{eq:B9}
            E^{TF}_{pol}=\int^{\zeta^{TF}_{pol}}_0\left[\zeta^2u(\zeta)+u^2(\zeta)\right]d\zeta ,
        \end{equation}
        and the energy for PM state in T-F limit,
        \begin{equation}
            E^{TF}_{PM}=\int^{\zeta^{TF}_{PM}}_0\left[\zeta^2u(\zeta)+(1+\lambda_1''/2)u^2(\zeta)\right]d\zeta.     
        \end{equation}
            
\end{document}